\documentclass[useAMS,usenatbib]{mn2e}
\usepackage[dvips]{graphicx}

\begin{document}

\title[Eccentric Ellipsoidal Variables in the LMC]{Eccentric Ellipsoidal Red Giant Binaries in the LMC: Complete Orbital Solutions and Comments on Interaction at Periastron}
\author[C. P. Nicholls \& P. R. Wood]{C. P. Nicholls$^{1,2}$\thanks{E-mail: cnicholls@physics.ucsd.edu (CPN); wood@mso.anu.edu.au (PRW)}
 and P. R. Wood$^{1}$\footnotemark[1]\\
$^{1}$Research School of Astronomy and Astrophysics, Australian National
University, Cotter Road, Weston Creek ACT 2611, Australia\\
$^{2}$Center for Astrophysics and Space Science, University of California San Diego, La Jolla, CA 92093, USA}

\date{Accepted 2012 January 4. Received 2011 December 13; in original form 2011 August 15.}

\pagerange{\pageref{firstpage}--\pageref{lastpage}} \pubyear{2012}

\maketitle

\label{firstpage}

\begin{abstract}

Modelling ellipsoidal variables with known distances can lead to exact determination of the masses of both components, even in the absence of eclipses. We present such modelling using light and radial velocity curves of ellipsoidal red giant binaries in the LMC, where they are also known as sequence E stars. Stars were selected as likely eccentric systems on the basis of light curve shape alone. We have confirmed their eccentric nature and obtained system parameters using the Wilson--Devinney code.

Most stars in our sample exhibit unequal light maxima as well as minima, a phenomenon not observed in sequence E variables with circular orbits. We find evidence that the shape of the red giant changes throughout the orbit due to the high eccentricity and the varying influence of the companion.

Brief intervals of pulsation are apparent in two of the red giants. We determine pulsation modes and comment on their placement in the period--luminosity plane.

Defining the parameters of these systems paves the way for modelling to determine by what mechanism eccentricity is maintained in evolved binaries.

\end{abstract}


\section{Introduction}
\label{intro}

Long period variables (LPVs) occupy several sequences in the period--luminosity plane \nocite{wood99mn} \nocite{ogle04} \citep[Wood et al.\ 1999; Soszy\'nski et al.\ 2004a;][]{ita04, fraser05, oglep-l, fraser08}. Most of these sequences represent radially pulsating stars on the red giant branch (RGB) or asymptotic giant branch (AGB), with different sequences corresponding to different modes of pulsation. There are two exceptions. Sequence D, lying at the longest periods, is occupied by AGB stars showing two concurrent forms of variation; these stars are also known as Long Secondary Period Variables, or LSPVs \citep[Wood et al.\ 1999;][]{hinkle02,sequenceDstars,seqDpaper,dust}. The other sequence lies close to sequence D and extends to the lowest luminosities. It was labelled by Wood et al.\ (1999) with the letter E and was shown to consist of red giant binaries. \nocite{ogleellipsoidal} Soszy\'nski et al. (2004b) identified the light variation as due to ellipsoidal variations, as alternate minima often have different depths. Sequence E stars exhibit regular light variations with periods between $\sim 50$ and 1000 days and amplitudes $\leq 0.3$ mag in the MACHO red band. They lie on both the RGB and the AGB.

Ellipsoidal variability is observed in close binaries, where the shape of a star is distorted by the gravitational influence of its companion. When the primary star evolves and begins to fill its Roche Lobe, it takes on an increasingly elongated, or `ellipsoidal', shape. Rotation of the star's aspherical shape causes variation in the light curve, even in the absence of eclipses. 

The hallmark of ellipsoidal variables is the relationship between the phased light and radial velocity curves. The radial velocity of the system is dominated by the orbital motion, but the light variability is mainly due to the change in apparent surface area of the distorted primary as it orbits its companion. Light maxima occur when the ellipsoidal star has one of its broad sides facing the observer and minima correspond to those times when the star is `end-on' from the observer's point of view. This orientation-induced light variability of the ellipsoidal star means the system's light curve displays two maxima and minima every orbit; the phased light curve shows two cycles for every cycle of the phased velocity curve.

Stars on sequence E were first unambiguously demonstrated to be ellipsoidal variables by \cite{seqEpaper}. In that paper we presented phased light and velocity curves of 11 sequence E binaries in the LMC, each showing doubling of the velocity period with respect to the light period (see figs.~1 and 2 of that paper). It is expected that the current red giant is the more massive star and the first to evolve, and that the less massive companion is normally on the main sequence (MS) and does not contribute significantly to the detected flux. 

A subset of the sequence E ellipsoidal red giant binaries was suggested by \nocite{ogleellipsoidal} Soszy\'nski et al. (2004b) to have eccentric orbits, based on their unusual light curve shapes. Eccentric orbits in close binaries with evolved components are unexpected, as tidal theory predicts that orbits should quickly circularise once stars begin to evolve \citep{zahn}. As an example, in \cite{seqEpaper} we calculated the circularisation time for sequence E stars using the formula given in \cite{soker00}. We found the typical circularisation time for sequence E binaries to be $\sim$ 3500 yr, much shorter than the lifetime of the ellipsoidal phase, which is $\sim 0.8$\,Myr \citep{nie11}. Eccentric orbits in evolved close binaries are thought to imply the presence of some mechanism that can maintain or increase eccentricity, opposing the tidal forces.

A number of other evolved binaries are also known to have unexpectedly eccentric orbits. These include post-AGB binaries \citep{vanwinckel03} and Barium stars \citep{rob10}. It is unclear why significantly nonzero orbital eccentricity is present in any of these systems, but several mechanisms have been suggested. These include mass transfer at periastron \citep{soker00} and tidal interaction with a circumbinary disk \citep{arty}.

The evolution of close binaries like ellipsoidal variables may end in a number of different ways, depending on the initial orbital separation and subsequent binary evolution. All observable ellipsoidal variables are partially filling their Roche Lobes. If the Roche Lobe is filled below the RGB tip, it is likely the binary will undergo a common envelope (CE) event and evolve slowly towards the white dwarf cooling track, as in this case the remnant star would be unable to heat up to and ionise the ejected stellar envelope before it disperses. If an ellipsoidal variable ends its evolution via the superwind at the AGB tip without filling its Roche Lobe, as single AGB stars do, it will produce a planetary nebula (PN) with a wide binary companion. Variables that fill their Roche Lobes somewhere on the AGB should undergo a CE event and become close binary PN. Further comment on the relationship of ellipsoidal variables to binary PN and asymmetric PN can be found in \cite{apn5paper}.

In this paper we analyse a sample of LMC sequence E binaries which display light curve shapes that \nocite{ogleellipsoidal} Soszy\'nski et al. (2004b) linked with eccentric orbits. We aim to confirm their eccentric nature and describe the components of each binary and their orbits, including estimates of absolute masses as is possible for ellipsoidal variables with known distances \citep[e.g.][]{wilson09}. The results provide input for future studies on the likelihood or otherwise of proposed mechanisms for maintaining eccentricity.
We use radial velocity curves obtained from observations taken with the ANU 2.3\,m telescope at Siding Spring Observatory (SSO) and OGLE II light curves. Modelling of the light and velocity curves is done with the 2010 version of the Wilson--Devinney code \citep{wd}.

\section{Observations and Data Reduction}

We selected a sample of seven sequence E stars with $I$-band magnitude brighter than 16 and light curves indicative of eccentric orbits from the OGLE II database. Table~\ref{startable} gives the OGLE identification and mean $V$ and $I$ magnitudes of each star. The stars were monitored using the Double Beam Spectrogaph \citep[DBS,][]{2.3m} at the ANU 2.3\,m telescope at SSO from May 2006 to April 2008, with the aim of obtaining radial velocities.

Our study used only the red arm of the DBS with a grating of 1200 lines/mm, a two-pixel resolution of 0.96\,\AA~and a grating angle of $32^{\circ}4'$, giving a wavelength range of approximately 8000 -- 9000\,\AA. This gave spectra centred on the Ca triplet, ideal for cross-correlation and the calculation of radial velocities. A Neon--Argon arc and an internal flat were taken after each exposure for calibration and to eliminate strong fringing on the CCDs. The spectra were taken over 15 runs throughout the monitoring period with 34 nights total observing. 

Data reduction was done in \textsc{iraf}. The spectra had the overscan bias subtracted and the overscan region removed from the images, and then the object and arc spectra were flatfielded and the spectra extracted using the \textit{apextract} package. The arcs were used to wavelength calibrate the object spectra, obvious cosmic rays were removed, and stars with multiple spectra taken on a single night were added using \textit{scombine}. Each star has between 13 and 17 spectra spread over the monitoring period.

Radial velocities were calculated using \textsc{iraf}'s cross-correlation task, \textit{fxcor}. A single spectrum with high signal to noise and narrow lines was selected from each star's collection to act as a template for that star's cross-correlation. Absolute radial velocities were provided by cross-correlation of the template with the radial velocity standard star $\alpha$ Cet, whose spectrum was taken on 11th November 2006 with the above observing configuration. Cross-correlation was done in the wavelength region 8370 -- 8920\,\AA, which was mostly free of telluric lines and included the Ca triplet. Spectra were then cross-correlated to a telluric spectrum in the region 8120 -- 8370\,\AA~(covering the bulk of the telluric lines) to check for zero point offsets in the reduction procedure. No zero point errors were found.

The OGLE $I$-band light curves and our new velocity curves of all 7 stars in the sample can be seen in Figs.~\ref{2013soln} --~\ref{3159soln}. The mean value of Heliocentric Julian Date (HJD) for each night's observations and the calculated radial velocities are given in Table~\ref{vtable}. Typical velocity errors are of the order of $2.5\,\rm{km\,s^{-1}}$ and are represented by the error bars in Figs.~\ref{2013soln} --~\ref{3159soln}.

\begin{table}
\centering
\caption[]{Candidate eccentric ellipsoidal variables in the OGLE II database. The OGLE ID contains each star's RA and Dec.}
\label{startable}
\begin{tabular}{lcc}
\hline
\multicolumn{1}{c}{OGLE ID}  &  \multicolumn{1}{c}{$V$}  &  \multicolumn{1}{c}{$I$}  \\
\hline                                                                           
OGLE052013.51-692253.2 & 16.62 & 15.17 \\
OGLE052438.40-700028.8 & 15.11 & 13.66 \\
OGLE052812.41-693417.9 & 17.33 & 15.78 \\
OGLE052850.12-701211.2 & 15.70 & 13.80 \\
OGLE053033.55-701742.0 & 15.34 & 13.87 \\
OGLE053124.49-701927.4 & 16.67 & 14.96 \\
OGLE053159.96-693439.5 & 16.01 & 14.29 \\
\hline
\end{tabular}
\end{table}

\begin{table*}
\centering
\caption[]{Radial velocities in $\rm{km\,s^{-1}}$ of eccentric sequence E stars. Stars are identified by their OGLE RA.}
\label{vtable}
\begin{tabular}{lccccccc}
\hline
\multicolumn{1}{c}{HJD}  &  \multicolumn{1}{c}{052013.51}  &  \multicolumn{1}{c}{052438.40}  &  \multicolumn{1}{c}{052812.41}  &  \multicolumn{1}{c}{052850.12}  &  \multicolumn{1}{c}{053033.55}  &  \multicolumn{1}{c}{053124.49}  &  \multicolumn{1}{c}{053159.96}  \\
\hline                                                                           
2453872.91  &  254.76  &       296.16   &       -      &     251.91  &     259.75   &       -      &       -     \\
2453981.20  &  253.11  &       283.78   &       -      &     238.42  &     237.13   &       -      &       -     \\
2453982.23  &     -      &          -       &       -      &        -      &        -       &    248.22  &    232.90 \\
2453983.23  &     -      &          -       &    261.03  &        -      &        -       &    246.62  &       -     \\
2454050.17  &  244.62  &       262.61   &       -      &     228.23  &     231.69   &    253.12  &       -     \\
2454051.10  &     -      &          -       &    239.52  &        -      &        -       &    251.42  &    227.26 \\
2454109.16  &     -      &       245.09   &       -      &     221.65  &     244.45   &       -      &       -     \\
2454110.04  &  231.78  &          -       &       -      &        -      &        -       &       -      &    232.08 \\
2454111.13  &     -      &          -       &    249.70  &        -      &        -       &    246.83  &       -     \\
2454136.22  &     -      &       245.40   &       -      &     227.83  &     257.28   &       -      &       -     \\
2454137.17  &  228.82  &          -       &       -      &        -      &        -       &    244.42  &    240.39 \\
2454137.99  &     -      &          -       &    255.28  &        -      &        -       &       -      &       -     \\
2454167.14  &     -      &       253.05   &       -      &        -      &        -       &       -      &       -     \\
2454167.98  &  239.62  &       247.33   &       -      &     232.92  &     268.22   &       -      &    254.69 \\
2454168.93  &     -      &          -       &    266.21  &        -      &        -       &    249.55  &       -     \\
2454206.89  &  243.56  &       265.81   &    263.00  &     241.77  &     271.55   &    250.76  &       -     \\
2454207.95  &     -      &          -       &       -      &        -      &        -       &       -      &    255.71 \\
2454310.25  &  261.01  &       297.26   &       -      &     254.92  &     249.89   &       -      &    250.59 \\
2454347.23  &     -      &          -       &       -      &     256.96  &     243.51   &    222.24  &    247.46 \\
2454348.31  &  253.91  &       294.11   &       -      &        -      &        -       &       -      &       -     \\
2454349.10  &     -      &          -       &       -      &        -      &        -       &       -      &       -     \\
2454378.17  &  261.02  &       288.66   &    260.25  &     257.33  &     237.30   &    230.38  &    243.08 \\
2454409.16  &  261.11  &       276.56   &       -      &     256.44  &     235.05   &    238.04  &    241.15 \\
2454452.23  &     -      &       265.40   &       -      &     256.67  &     235.23   &    247.99  &    233.66 \\
2454453.95  &     -      &          -       &    275.04  &        -      &        -       &       -      &       -     \\
2454454.10  &     -      &          -       &    267.86  &        -      &        -       &       -      &       -     \\
2454524.07  &     -      &       243.97   &       -      &        -      &        -       &       -      &       -     \\
2454543.10  &  241.48  &       242.95   &       -      &     250.30  &     264.18   &    255.56  &    228.90 \\
2454543.94  &     -      &          -       &    245.65  &        -      &        -       &       -      &       -     \\
2454545.09  &  234.34  &          -       &    247.01  &        -      &        -       &    253.69  &    231.90 \\
2454571.07  &  215.03  &       244.55   &       -      &     248.54  &     270.34   &    261.19  &    228.57 \\
2454571.87  &  223.48  &          -       &    236.37  &        -      &        -       &       -      &       -     \\
2454573.00  &  221.59  &       245.22   &       -      &     249.44  &        -       &       -      &       -     \\
\hline
\end{tabular}
\end{table*}

\section{Absolute Solutions of Orbital and Stellar Parameters}

Analysing both the light and velocity curves of ellipsoidal variables at known distances provides a unique opportunity to determine complete binary solutions, even in the absence of eclipses. Implementation of this within the Wilson--Devinney (WD) code is known as Inverse Distance Estimation (IDE) \citep{wilson09}. From a conceptual perspective, the basic method is as follows. If the distance, observed magnitude and extinction are known, the absolute luminosity of an ellipsoidal variable can be derived. If the temperature is also known (e.g. from spectra or colour), the stellar radius can be calculated.

If the orbital inclination of an ellipsoidal variable is known or can be constrained, then the amplitude of the light variation measures what fraction of its Roche Lobe the ellipsoidal star fills. The previously calculated radius therefore gives the size of the Roche Lobe, $R_{\rm{lobe}}$. The radial velocity curve provides the semimajor axis of the red giant's orbit, $a_{1}$, so a solution for mass ratio $q$ is possible, since $q$ is a function of $R_{\rm{lobe}}$/$a_{1}$. So for an assumed $i$, the absolute masses of stars are produced. Therefore we need step only the inclination in our solutions to find the best fit to the light and velocity curves.

Our analysis of ellipsoidal red giant binaries with undetectable companions and known distances has been preceded by analysis of a similar binary by \cite{wilson09}, who also pioneered the method described above and conveniently added this capability to the 2010 version of the WD code. The reader is referred to that paper for a more in-depth explanation of absolute light and velocity analysis of ellipsoidal variables.

\subsection{Modelling the Light and Velocity Curves}
\label{model}

To describe our ellipsoidal variables and obtain their orbital parameters, we used the 2010 version of the WD code \citep{wd,wilsonDDE,wilson09}. Preliminary fits to the velocity curves were made with a Fortran program, \textsc{Fitall}, to obtain starting estimates of input parameters for the WD modelling. This allowed us to fix the system velocity $v_{\gamma}$, and gave starting estimates for the semimajor axis $a$, angle of periastron $\omega$, eccentricity $e$, and mass ratio $q$. Some of these initial parameters are shown in Table~\ref{paramtable}.

The luminosity of each star was calculated using its median OGLE $V-I$ colour, a bolometric correction to $I$ calculated from the \cite{houdashelt} models for K and M giants, the LMC distance modulus 18.54 and reddening $E(B-V) = 0.08$ \citep{kellerwood}, and $V$ and $I$ extinction calculated using the \cite{cardelli} equations. Effective temperatures $T_{\rm{eff}}$ were calculated from a fit to the $(T_{\rm{eff}}, V-I)$ data of \cite{houdashelt}, and stellar radii were calculated from luminosity and $T_{\rm{eff}}$ using the Stefan--Boltzmann law. The temperature of the ellipsoidal red giant $T_{\rm{eff_1}}$, an input for both the \textsc{lc} and \textsc{dc} programs of the WD code, was fixed from these calculations. The WD code uses bolometric corrections and bandpass fluxes computed from \cite{kurucz} model atmospheres. After fitting the observed curves, and using the known distance, the code calculates the stellar luminosity. This agreed well with the value computed with the \cite{houdashelt} bolometric corrections, showing consistency of the model atmospheres.

A metallicity of $[M/H] = -0.3$ was used for these LMC stars. Zero point flux calibrations for the $I_c$ band (OGLE light curve) and the Johnson $I$ band (the best match to the wavelengths of the observed spectra from which velocity curves were calculated) were taken from \cite{bessell79} and \cite{johnson66}, respectively. All stars were assumed to have their rotation velocities periastron-synchronised and the rotation parameters $F_1$ and $F_2$ were calculated accordingly. The invisible companion star to each ellipsoidal red giant was initially assumed to be a sunlike MS star with $T_{\rm{eff_2}} = 6000\,K$, and the surface potentials were set so that the companions made no contribution to the modelled light curves. The parameters of the companion were later adjusted as described below for each case. The potentials of the red giants were set during initial explorations with the \textsc{lc} program so the modelled luminosities and radii matched the independently calculated values.

Solutions for all stars were performed in mode 2, for detached binaries. We used simple reflection treatment with no spots, and no proximity effects. Limb darkening was done via the square root law with coefficients calculated locally from the \cite{vanhamme93} tables. The stellar atmosphere formulation was used for local flux emission calculations instead of the less accurate blackbody formulation. The gravity darkening exponents were set to 0.3 for each red giant and 1.0 for each MS companion, as appropriate for convective and radiative envelopes, respectively. Bolometric albedos were set to 0.5 for each red giant and 1.0 for each companion, again as expected for convective and radiative envelopes. Symmetric derivatives were used in all solutions to improve convergence.

The inclination was stepped down from $90^{\circ}$, in $10^{\circ}$ increments, producing a one-dimensional family of solutions for each system. Stepping of $i$ stopped when either $q$ exceeded unity or the solution was deemed too poor. We allowed \textsc{dc} to iterate on $a$, $e$, $\omega$, primary star potential $\Omega_1$, and $q$. The sum of squares of residuals of the light curve was noted for each solution, as a means of quantitatively determining the best solution. The velocity curve residuals were also examined but were found to not provide useful additional constraints on orbital inclination, as they varied randomly or without minima. In contrast to the light curves, the velocity curves have fewer features and are less well sampled. Any change in velocity amplitude due to inclination can be compensated for in the models by changing the stellar masses, so we did not include velocity residuals in our estimation of the best solution. In general, solutions converged (so that corrections $\ll$ errors) in 6 to 8 iterations.

The phase of the modelled light and velocity curves was calculated by \textsc{lc}, using the input phase zero point, $\rm{HJD_0}$. The convention in binary analysis is for superior conjunction to occur at or near phase zero. To find $\rm{HJD_0}$ for each star we noted the light curve shape at the phase of superior conjunction according to the velocity curve, then visually inspected the light curve plotted against Julian Date to find the HJD of superior conjunction, choosing the earliest value in the data. The final value of $\rm{HJD_0}$ was obtained through tweaking of the phased plots to find a good match between the theoretical and observed curves. We calculated the phasing of the observed curves using the same value of $\rm{HJD_0}$ as for the theoretical curves. The values of $\rm{HJD_0}$ for each star are shown in Table~\ref{paramtable}.

\begin{table}
\centering
\caption[]{Initial parameters for our eccentric sequence E sample. Stars are identified by their OGLE RA.}
\label{paramtable}
\begin{tabular}{lcccc}
\hline
\multicolumn{1}{c}{Star}  &  \multicolumn{1}{c}{$\rm{HJD_0}$}  &  \multicolumn{1}{c}{$P$ (days)}  &  \multicolumn{1}{c}{$L (L_{\odot})$}  &  \multicolumn{1}{c}{$T_{\rm{eff}}$ (K)}  \\
\hline                                                                           
052013.51 & 2450390.0 & 452.47 & 1196.16 &  4221.71\\
052438.40 & 2450315.0 & 410.96 & 4747.84 &  4240.74\\
052812.41 & 2450640.0 & 258.70 & 721.33  & 4064.10 \\
052850.12 & 2449955.0 & 662.20 & 5441.67 &  3718.76\\
053033.55 & 2450800.0 & 390.17 & 4017.29 &  4183.04\\
053124.49 & 2450440.0 & 541.32 & 1671.59 &  3877.35\\
053159.96 & 2450870.0 & 501.10 & 3130.32 &  3876.33\\
\hline
\end{tabular}
\end{table}

\subsection{Individual star solutions}
\label{solns}

\subsubsection{OGLE052013.51}

\begin{figure}
\begin{center}
\includegraphics[width=0.5\textwidth]{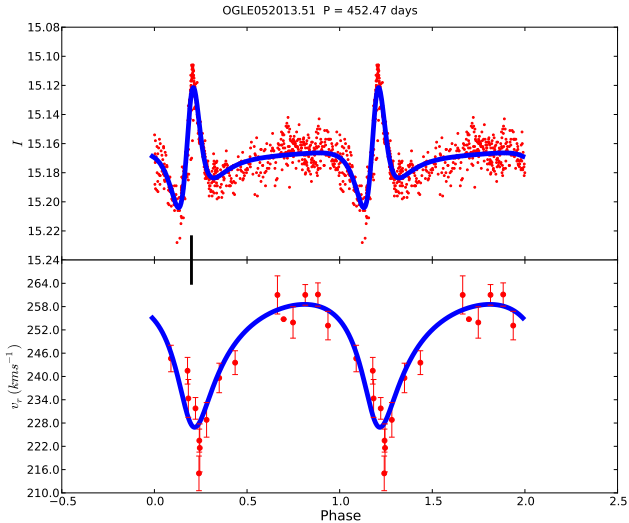}
\end{center}
\caption[]{Observed OGLE $I$ light curve and observed radial velocity curve for OGLE052013.51 (red points) and modelled light and velocity curves (blue lines) at an inclination of $60^{\circ}$. The vertical black line marks the phase of periastron.}
\label{2013soln}
\end{figure}

The observed light and velocity curves are shown in Fig.~\ref{2013soln}. The light curve has unequal maxima and minima, with one of the maxima very sharp. At superior conjunction (when the red giant is behind its companion from an observer's point of view), the corresponding minimum of the light curve is deeper, meaning the red giant is dimmer on the end closest its companion. As explained in \cite{seqEpaper}, this is likely due to gravity darkening.

The angle of periastron is $\sim 160^{\circ}$, meaning periastron occurs between superior conjunction and the following maximum, at a phase of $\sim 0.2$. This and the high orbital eccentricity ($\sim 0.4$) explains why the brighter maximum is so sharp: in a highly eccentric orbit, a star moves very fast at periastron and so that part of the light curve spans a shorter time interval. This however cannot explain why the star is brighter there than at its other maximum. A hypothesis to explain this brightening is presented in Section \ref{distortion}.

The results of our modelling for various $i$ values are shown in Table~\ref{2013table}, where $\Sigma r^{2} (I)$ denotes the sum of squares of residuals of the $I$-band light curve from the modelled curve. The errors shown are the standard errors calculated by \textsc{dc} for each adjusted parameter. Semimajor axis, $e$, $\omega$, $\Omega_1$ and $q$ all increase significantly with decreasing $i$. The mass ratio exceeded unity at $i=40^{\circ}$ so solutions were not made at lower inclinations.

According to the sum of squares of residuals of the light curve, the best solution for a compact secondary is at $i=60^{\circ}$. This solution was further refined by obtaining more accurate estimates of the MS companion's properties. At $i = 60^{\circ}$, the red giant has a mass of $1.4\,M_{\odot}$, and a radius of $66.3\,R_{\odot}$. From the colour temperature of 4220\,K, we calculate the luminosity as $1252\,L_{\odot}$. Using the evolutionary track data of \cite{girardi}, a red giant of this mass and luminosity and of LMC metallicity should have an age of $\sim 3 \times 10^{9}$ y and be on the RGB. At the same age, the $1.1\,M_{\odot}$ MS companion should have $L = 2.4\,L_{\odot}$, $T_{\rm{eff}} = 6490\,K$, and $R = 1.2\,R_{\odot}$. We re-solved the system at $i = 60^{\circ}$ with $T_{\rm{eff_2}}$ increased to 6490\,K and the companion's surface potential, $\Omega_2$, altered so $R_2 = 1.2\,R_{\odot}$. This more accurate solution is shown in the last row of Table~\ref{2013table} and in Fig.~\ref{2013soln}. Increasing the accuracy of the companion parameters caused no significant changes in the solution.

\begin{table*}
\centering
\caption[]{Simultaneous light and velocity solutions for OGLE052013.51. \dag \ denotes the final solution at the best inclination and with the most accurate companion properties.}
\label{2013table}
\begin{tabular}{cccccccccc}
\hline
\multicolumn{1}{c}{$i$} & \multicolumn{1}{c}{$a (\rm{R_{\odot}})$} & \multicolumn{1}{c}{$e$} & \multicolumn{1}{c}{$\omega$ (radians)} & \multicolumn{1}{c}{$\Omega_{1}$} & \multicolumn{1}{c}{$q$} & \multicolumn{1}{c}{$M_{1} (\rm{M_{\odot}})$} & \multicolumn{1}{c}{$M_{2} (\rm{M_{\odot}})$} & \multicolumn{1}{c}{$R_{1} (\rm{R_{\odot}})$} & \multicolumn{1}{c}{$\Sigma r^{2} (I)$}  \\
\hline     
$90^{\circ}$ & $331.596 \pm 6.065$ & $0.393 \pm 0.005$ & $2.789 \pm 0.007$  & $6.180 \pm 0.165$ & $0.573 \pm 0.045$ & 1.52 & 0.87 & 68.61 & $0.630 \times 10^{-15}$ \\
$80^{\circ}$ & $331.749 \pm 5.937$ & $0.396 \pm 0.005$ & $2.791 \pm 0.007$  & $6.243 \pm 0.162$ & $0.594 \pm 0.045$ & 1.50 & 0.89 & 68.33 & $0.616 \times 10^{-15}$ \\
$70^{\circ}$ & $331.820 \pm 5.603$ & $0.402 \pm 0.005$ & $2.797 \pm 0.006$  & $6.422 \pm 0.158$ & $0.655 \pm 0.046$ & 1.45 & 0.95 & 67.53 & $0.583 \times 10^{-15}$ \\
$60^{\circ}$ & $336.724 \pm 5.194$ & $0.416 \pm 0.005$ & $2.809 \pm 0.006$  & $6.836 \pm 0.155$ & $0.782 \pm 0.047$ & 1.41 & 1.10 & 66.32 & $0.563 \times 10^{-15}$ \\
$50^{\circ}$ & $348.620 \pm 5.291$ & $0.442 \pm 0.006$ & $2.830 \pm 0.006$  & $7.478 \pm 0.166$ & $0.949 \pm 0.052$ & 1.43 & 1.35 & 65.01 & $0.663 \times 10^{-15}$ \\
$40^{\circ}$ & $372.829 \pm 6.981$ & $0.498 \pm 0.009$ & $2.859 \pm 0.006$  & $8.222 \pm 0.212$ & $1.023 \pm 0.063$ & 1.68 & 1.72 & 64.13 & $0.866 \times 10^{-15}$ \\
\dag \ $60^{\circ}$ & $337.312 \pm 5.130$ & $0.416 \pm 0.005$ & $2.809 \pm 0.006$  & $6.858 \pm 0.154$ & $0.789 \pm 0.046$ & 1.41 & 1.11 & 66.29 & $0.566 \times 10^{-15}$ \\
\hline
\end{tabular}
\end{table*}

\subsubsection{OGLE052438.40}

\begin{figure}
\begin{center}
\includegraphics[width=0.5\textwidth]{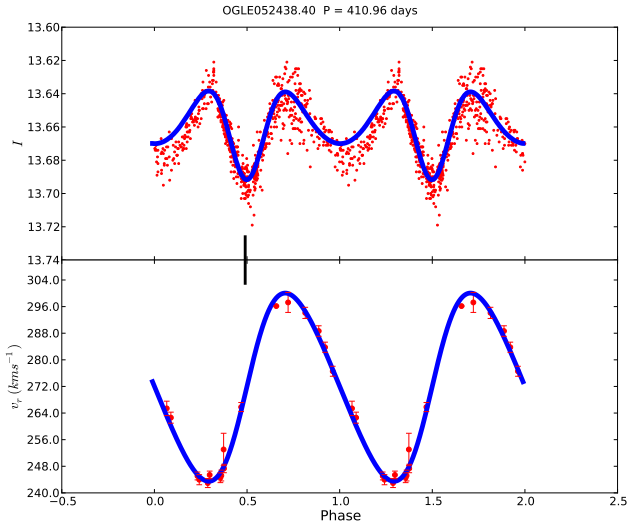}
\end{center}
\caption[]{Observed OGLE $I$ light curve and observed radial velocity curve for OGLE052438.40 (red points) and modelled light and velocity curves (blue lines) at an inclination of $70^{\circ}$. The vertical black line marks the phase of periastron.}
\label{2438soln}
\end{figure}

From Fig.~\ref{2438soln} it is clear that this variable displays equal maxima and unequal minima. However, the most notable property of this star's light curve is that the deeper minimum occurs at a different place with respect to the velocity curve than expected. Instead of being dimmer towards the companion as gravity darkening dictates, the red giant is dimmer at its other light minimum (inferior conjunction, or when the `outer end' of the red giant is towards us). An explanation for this is proposed in Section \ref{distortion}

The angle of periastron is $\sim 268^{\circ}$, so periastron occurs about the same time as inferior conjunction, or at a phase of $\sim 0.5$.

The results of our modelling with different $i$ values can be found in Table~\ref{2438table}. For most solutions, $a$, $e$ and $\omega$ are not significantly different. However $\Omega_1$ and $q$ change significantly with inclination.

According to the sum of squares of residuals of the light curve, the best solution is at $90^{\circ}$. However using the \cite{girardi} evolutionary tracks as above to find the temperature and radius of the MS companion resulted in eclipses in the light curve at both $90^{\circ}$ and $80^{\circ}$ inclinations. The eclipses disappeared at $i = 70^{\circ}$, where the red giant has a mass of $5.8\,M_{\odot}$, and a radius of $131\,R_{\odot}$. From the temperature of 4240\,K, we calculate the luminosity as $4979\,L_{\odot}$. A red giant of this mass and luminosity and at LMC metallicity should have an age of $\sim 7 \times 10^{7}$ y, and lie on the RGB. At the same age, the $5\,M_{\odot}$ MS companion should have $L = 881\,L_{\odot}$, T = 16710\,K, and $R = 3.5\,R_{\odot}$. We re-solved the system at $i = 70^{\circ}$ with $T_{\rm{eff _2}}$ increased to 16710\,K and $\Omega_2$ altered so $R_2 = 3.5\,R_{\odot}$. This solution is shown in the second last row of Table~\ref{2438table} and in Fig.~\ref{2438soln}. The only significant change in parameters for an accurately sized secondary is a slightly higher $\omega$. 

Both the stars in this system are of higher mass than the general LMC intermediate mass population. In particular, this system has the highest mass ratio of our sample, and the secondary has the highest mass, temperature, and luminosity and the greatest radius of the modelled companions. Therefore OGLE052438.40 is the system most likely to show a contribution of the MS companion to the overall flux. Using the derived luminosities, the secondary contributes 0.07 mag. to the flux at $V$ and 0.02 mag. to the flux at $I$, meaning the red giant's $V-I$ colour should reduce by 0.05, corresponding to a $\sim 50$\,K drop in the red giant's estimated temperature. To check whether this made a significant change to the modelled solution, we re-solved the system at the best inclination ($70^{\circ}$), with $T_{\rm{eff _1}}$ reduced to 4190\,K and the initial value of $\Omega_1$ altered so $R_1$ was increased to $134\,R_{\odot}$ (to account for the radius increase factor of 1.024 associated with the temperature change). This refined solution is shown in the last row of Table~\ref{2438table}. The most notable differences are in the semimajor axis and the masses, all of which have increased significantly. However there was no discernible change in the shape of the modelled light curve. The contribution of the secondary to the system light and orbital solutions was insignificant in all other systems studied here.

\begin{table*}
\centering
\caption[]{Simultaneous light and velocity solutions for OGLE052438.40. \dag \ denotes the solution at the best inclination and with the most accurate companion properties. \ddag \ denotes the final solution including the light contribution from the secondary.}
\label{2438table}
\begin{tabular}{cccccccccc}
\hline
\multicolumn{1}{c}{$i$}  &  \multicolumn{1}{c}{$a (\rm{R_{\odot}})$}  &  \multicolumn{1}{c}{$e$}  &  \multicolumn{1}{c}{$\omega$ (radians)}  &  \multicolumn{1}{c}{$\Omega_{1}$}  &  \multicolumn{1}{c}{$q$}  &  \multicolumn{1}{c}{$M_{1}( \rm{M_{\odot}})$}  &  \multicolumn{1}{c}{$M_{2} (\rm{M_{\odot}})$}  &  \multicolumn{1}{c}{$R_{1} (\rm{R_{\odot}})$} & \multicolumn{1}{c}{$\Sigma r^{2} (I)$}  \\
\hline     
$90^{\circ}$  & $515.379 \pm 4.270$  & $0.135 \pm 0.006$  & $4.640 \pm 0.033$  & $4.880 \pm 0.050$  & $0.770 \pm 0.019$  & 6.15  & 4.74  & 131.41  & $0.699 \times 10^{-14}$ \\
$80^{\circ}$  & $514.782 \pm 4.269$  & $0.135 \pm 0.006$  & $4.633 \pm 0.033$  & $4.904 \pm 0.050$  & $0.791 \pm 0.019$  & 6.06  & 4.79  & 131.30  & $0.700 \times 10^{-14}$ \\
$70^{\circ}$  & $513.360 \pm 4.189$  & $0.134 \pm 0.006$  & $4.635 \pm 0.033$  & $4.997 \pm 0.052$  & $0.869 \pm 0.021$  & 5.76  & 5.00  & 130.96  & $0.700 \times 10^{-14}$ \\
$60^{\circ}$  & $511.935 \pm 4.319$  & $0.130 \pm 0.006$  & $4.631 \pm 0.035$  & $5.186 \pm 0.058$  & $1.025 \pm 0.026$  & 5.27  & 5.40  & 130.35  & $0.703 \times 10^{-14}$ \\
$50^{\circ}$  & $511.608 \pm 4.446$  & $0.122 \pm 0.005$  & $4.624 \pm 0.039$  & $5.576 \pm 0.070$  & $1.342 \pm 0.036$  & 4.55  & 6.10  & 129.36  & $0.715 \times 10^{-14}$ \\
\dag \ $70^{\circ}$  & $511.549 \pm 4.994$  & $0.138 \pm 0.006$  & $4.676 \pm 0.034$  & $5.064 \pm 0.074$  & $0.899 \pm 0.031$  & 5.61  & 5.04  & 129.55  & $0.543 \times 10^{-14}$ \\
\ddag \ $70^{\circ}$  & $521.856 \pm 5.078$  & $0.139 \pm 0.006$  & $4.677 \pm 0.034$  & $4.979 \pm 0.072$  & $0.867 \pm 0.030$  & 6.06  & 5.25  & 133.85  & $0.717 \times 10^{-14}$ \\
\hline
\end{tabular}
\end{table*}

\subsubsection{OGLE052812.41}

\begin{figure}
\begin{center}
\includegraphics[width=0.5\textwidth]{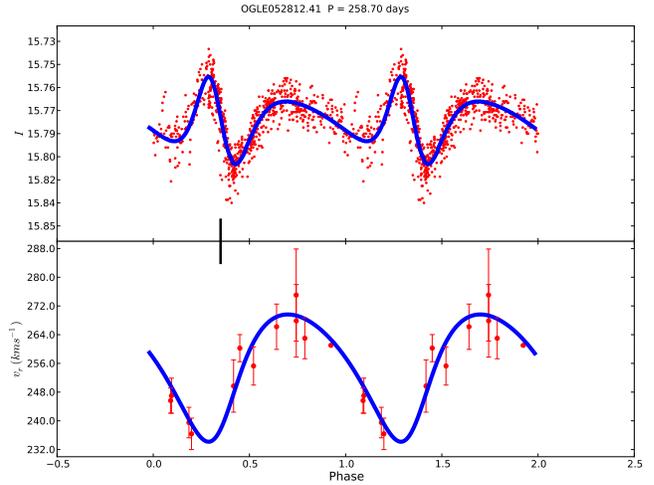}
\end{center}
\caption[]{Observed OGLE $I$ light curve and observed radial velocity curve for OGLE052812.41 (red points) and modelled light and velocity curves (blue lines) at an inclination of $90^{\circ}$. The vertical black line marks the phase of periastron.}
\label{2812soln}
\end{figure}

The observed light and velocity curves of this star are depicted in Fig.~\ref{2812soln}. Its light curve has unequal maxima and minima, and again the deeper minimum of the light curve occurs at inferior conjunction, instead of at superior conjunction as expected for an ellipsoidal variable. 

The angle of periastron of the red giant is $\sim 215^{\circ}$, which means periastron occurs just after the brighter maximum of the light curve, before inferior conjunction, at a phase of $\sim 0.35$. Again, proximity of the narrower maximum to periastron is explained by the eccentricity of the orbit.

The solutions for different inclinations are shown in Table~\ref{2812table}. For most $i$, $a$ and $e$ do not differ significantly. Angle of periastron and $\Omega_1$ vary slighty with $i$ while $q$ varies more significantly. Mass ratio exceeded unity at $i=50^{\circ}$.

The best solution for a small secondary star is at $90^{\circ}$, according to the sum of squares of residuals of the light curve. This solution was refined further by obtaining more accurate estimates of the MS companion's properties. At $i = 90^{\circ}$, the red giant has a mass of $1.4\,M_{\odot}$, and a radius of $55.7\,R_{\odot}$. With a temperature of 4060\,K, the luminosity is $756\,L_{\odot}$. Using the data of \cite{girardi}, a red giant of this mass and luminosity at LMC metallicity should have an age of $\sim 3 \times 10^{9}$ y, and lie on the RGB. At the same age, the $0.9\,M_{\odot}$ MS companion should have $L = 0.7\,L_{\odot}$, T = 5790\,K, and $R = 0.8\,R_{\odot}$. We further refined our best solution ($i = 90^{\circ}$) with $T_{\rm{eff _2}}$ decreased to 5790\,K and $\Omega_2$ altered so $R_2 = 0.8\,R_{\odot}$. This solution is shown in the last row of Table~\ref{2812table} and in Fig.~\ref{2812soln}, and is almost unchanged from the original solution with $i = 90^{\circ}$.

\begin{table*}
\centering
\caption[]{Simultaneous light and velocity solutions for OGLE052812.41. \dag \ denotes the final solution at the best inclination and with the most accurate companion properties.}
\label{2812table}
\begin{tabular}{cccccccccc}
\hline
\multicolumn{1}{c}{$i$}  &  \multicolumn{1}{c}{$a (\rm{R_{\odot}})$}  &  \multicolumn{1}{c}{$e$}  &  \multicolumn{1}{c}{$\omega$ (radians)}  &  \multicolumn{1}{c}{$\Omega_{1}$}  &  \multicolumn{1}{c}{$q$}  &  \multicolumn{1}{c}{$M_{1} (\rm{M_{\odot}})$}  &  \multicolumn{1}{c}{$M_{2} (\rm{M_{\odot}})$}  &  \multicolumn{1}{c}{$R_{1} (\rm{R_{\odot}})$} & \multicolumn{1}{c}{$\Sigma r^{2} (I)$}  \\
\hline     
$90^{\circ}$ & $225.412 \pm 4.920$  & $0.237 \pm 0.007$  & $3.758 \pm 0.022$  & $4.985 \pm 0.145$  & $0.652 \pm 0.045$  & 1.39 &  0.91 &  55.65 & $0.182 \times 10^{-15}$ \\
$80^{\circ}$ & $225.634 \pm 4.906$  & $0.236 \pm 0.007$  & $3.752 \pm 0.022$  & $5.025 \pm 0.147$  & $0.675 \pm 0.047$  & 1.38 &  0.93 &  55.60 & $0.183 \times 10^{-15}$ \\
$70^{\circ}$ & $226.185 \pm 4.633$  & $0.236 \pm 0.007$  & $3.741 \pm 0.022$  & $5.148 \pm 0.145$  & $0.750 \pm 0.049$  & 1.33 &  1.00 &  55.44 & $0.184 \times 10^{-15}$ \\
$60^{\circ}$ & $223.762 \pm 4.280$  & $0.234 \pm 0.007$  & $3.712 \pm 0.023$  & $5.257 \pm 0.141$  & $0.849 \pm 0.052$  & 1.22 &  1.03 &  55.17 & $0.191 \times 10^{-15}$ \\
$50^{\circ}$ & $221.307 \pm 4.066$  & $0.225 \pm 0.008$  & $3.663 \pm 0.025$  & $5.466 \pm 0.145$  & $1.023 \pm 0.059$  & 1.08 &  1.10 &  54.76 & $0.215 \times 10^{-15}$ \\
\dag \ $90^{\circ}$ & $224.800 \pm 4.926$  & $0.236 \pm 0.007$  & $3.755 \pm 0.022$  & $4.965 \pm 0.144$  & $0.644 \pm 0.045$  & 1.39 &  0.89 &  55.64 & $0.183 \times 10^{-15}$ \\
\hline
\end{tabular}
\end{table*}

\subsubsection{OGLE052850.12}

\begin{figure}
\begin{center}
\includegraphics[width=0.5\textwidth]{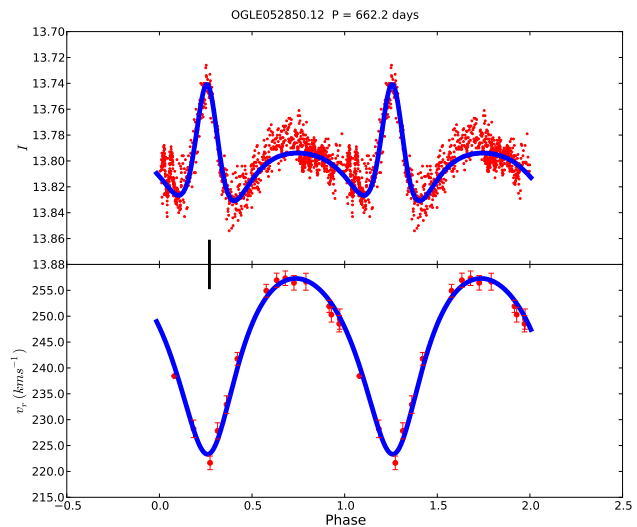}
\end{center}
\caption[]{Observed OGLE $I$ light curve and observed radial velocity curve for OGLE052850.12 (red points) and modelled light and velocity curves (blue lines) at an inclination of $50^{\circ}$. The vertical black line marks the phase of periastron.}
\label{2850soln}
\end{figure}

The observed light and radial velocity curves are shown in Fig.~\ref{2850soln}. The light curve shows unequal maxima and minima of almost equal depths.

The angle of periastron is $\sim 188^{\circ}$, so periastron occurs almost concurrently with the brighter light maximum, at a phase of $\sim 0.27$. The narrower maximum is again due to the star moving quickly at periastron during this part of its orbit.

The results are shown in Table~\ref{2850table}. Semimajor axis, $e$, $\omega$, and $\Omega_1$ vary slowly with $i$, while $q$ changes rapidly with $i$, particularly at low inclinations, and exceeds unity at $i=40^{\circ}$.

According to the sum of the squares of residuals of the light curve, the best solution is at $i=50^{\circ}$. This solution was further refined by obtaining more accurate estimates of the MS companion's properties. At $i = 50^{\circ}$, the red giant has a mass of $4.3\,M_{\odot}$, and a radius of $176\,R_{\odot}$. From the colour temperature of 3720\,K, we calculate the luminosity as $5325\,L_{\odot}$. A red giant of this mass and luminosity and at LMC metallicity should have an age of $\sim 1.5 \times 10^{8}$ y and be on the AGB. At the same age, the $3.4\,M_{\odot}$ MS companion should have $L = 242\,L_{\odot}$, T = 13430\,K, and $R = 2.88\,R_{\odot}$. We re-solved this solution with $T_{\rm{eff _2}}$ increased to 13430\,K and $\Omega_2$ altered so $R_2 = 2.88\,R_{\odot}$. This solution is shown in the last row of Table~\ref{2850table} and in Fig.~\ref{2850soln}.

\begin{table*}
\centering
\caption[]{Simultaneous light and velocity solutions for OGLE052850.12. \dag \ denotes the final solution at the best inclination and with the most accurate companion properties.}
\label{2850table}
\begin{tabular}{cccccccccc}
\hline
\multicolumn{1}{c}{$i$}  &  \multicolumn{1}{c}{$a (\rm{R_{\odot}})$}  &  \multicolumn{1}{c}{$e$}  &  \multicolumn{1}{c}{$\omega$ (radians)}  &  \multicolumn{1}{c}{$\Omega_{1}$}  &  \multicolumn{1}{c}{$q$}  &  \multicolumn{1}{c}{$M_{1} (\rm{M_{\odot}})$}  &  \multicolumn{1}{c}{$M_{2} (\rm{M_{\odot}})$}  &  \multicolumn{1}{c}{$R_{1} (\rm{R_{\odot}})$} & \multicolumn{1}{c}{$\Sigma r^{2} (I)$}  \\
\hline     
$90^{\circ}$  & $648.261 \pm 6.646$  & $0.241 \pm 0.005$  & $3.368 \pm 0.021$  & $4.302 \pm 0.056$  & $0.466 \pm 0.016$  & 5.69  & 2.65  & 182.79  & $0.201 \times 10^{-13}$ \\
$80^{\circ}$  & $649.654 \pm 5.285$  & $0.244 \pm 0.005$  & $3.359 \pm 0.020$  & $4.352 \pm 0.044$  & $0.489 \pm 0.013$  & 5.64  & 2.76  & 182.38  & $0.198 \times 10^{-13}$ \\
$70^{\circ}$  & $642.750 \pm 4.315$  & $0.249 \pm 0.005$  & $3.355 \pm 0.018$  & $4.416 \pm 0.036$  & $0.540 \pm 0.011$  & 5.28  & 2.85  & 181.26  & $0.185 \times 10^{-13}$ \\
$60^{\circ}$  & $632.234 \pm 6.562$  & $0.255 \pm 0.005$  & $3.326 \pm 0.016$  & $4.526 \pm 0.065$  & $0.623 \pm 0.022$  & 4.77  & 2.97  & 179.32  & $0.167 \times 10^{-13}$ \\
$50^{\circ}$  & $630.260 \pm 4.228$  & $0.254 \pm 0.005$  & $3.286 \pm 0.015$  & $4.821 \pm 0.045$  & $0.796 \pm 0.018$  & 4.27  & 3.40  & 176.43  & $0.155 \times 10^{-13}$ \\
$40^{\circ}$  & $639.198 \pm 5.137$  & $0.228 \pm 0.006$  & $3.256 \pm 0.019$  & $5.350 \pm 0.063$  & $1.114 \pm 0.030$  & 3.79  & 4.22  & 172.54  & $0.196 \times 10^{-13}$ \\
\dag \ $50^{\circ}$  & $628.582 \pm 5.807$  & $0.252 \pm 0.005$  & $3.287 \pm 0.015$  & $4.835 \pm 0.068$  & $0.807 \pm 0.027$  & 4.21  & 3.40  & 175.93  & $0.155 \times 10^{-13}$ \\
\hline     
\end{tabular}
\end{table*}

\subsubsection{OGLE053033.55}

\begin{figure}
\begin{center}
\includegraphics[width=0.5\textwidth]{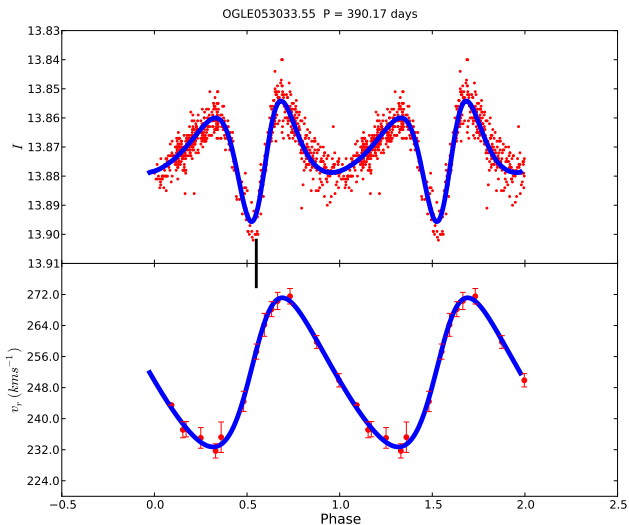}
\end{center}
\caption[]{Observed OGLE $I$ light curve and observed radial velocity curve for OGLE053033.55 (red points) and modelled light and velocity curves (blue lines) at an inclination of $60^{\circ}$. The vertical black line marks the phase of periastron.}
\label{3033soln}
\end{figure}

The observed light and velocity curves are depicted in Fig.~\ref{3033soln}. This star also shows unequal maxima and minima, with the deeper minimum of the light curve at inferior conjunction.

The angle of periastron is $\sim 290^{\circ}$, with periastron occuring between inferior conjunction and the following light maximum, at a phase of $\sim 0.55$. 

The solutions for all modelled inclinations and a compact secondary are shown in Table~\ref{3033table}. Semimajor axis, $e$, $\omega$, $\Omega_1$ and $q$ all vary significantly with $i$. The mass ratio exceeded unity at $i=30^{\circ}$, and we did not make solutions at lower inclinations.

According to the sum of squares of residuals of the light curve, the best solution is at $60^{\circ}$. We further refined this solution by obtaining more accurate estimates of the MS companion's properties. At $i = 60^{\circ}$, the red giant has a mass of $5.8\,M_{\odot}$, and a radius of $123\,R_{\odot}$. The colour temperature of 4180\,K gives a luminosity of $4147\,L_{\odot}$. Using the data of \cite{girardi}, a red giant of this mass and luminosity and at LMC metallicity should have an age of $\sim 7 \times 10^{7}$ y, and lie on the RGB. At the same age, the $3.2\,M_{\odot}$ MS companion should have $L = 99\,L_{\odot}$, T = 13310\,K, and $R = 1.9\,R_{\odot}$. We re-solved at $i = 60^{\circ}$ with $T_{\rm{eff _2}}$ increased to 13310\,K and $\Omega_2$ altered so $R_2 = 1.9\,R_{\odot}$. This solution is shown in the last row of Table~\ref{3033table} and in Fig.~\ref{3033soln}. The masses in this system are again higher than the typical LMC field population.

\begin{table*}
\centering
\caption[]{Simultaneous light and velocity solutions for OGLE053033.55. \dag \ denotes the final solution at the best inclination and with the most accurate companion properties.}
\label{3033table}
\begin{tabular}{cccccccccc}
\hline
\multicolumn{1}{c}{$i$}  &  \multicolumn{1}{c}{$a (\rm{R_{\odot}})$}  &  \multicolumn{1}{c}{$e$}  &  \multicolumn{1}{c}{$\omega$ (radians)}  &  \multicolumn{1}{c}{$\Omega_{1}$}  &  \multicolumn{1}{c}{$q$}  &  \multicolumn{1}{c}{$M_{1} (\rm{M_{\odot}})$}  &  \multicolumn{1}{c}{$M_{2} (\rm{M_{\odot}})$}  &  \multicolumn{1}{c}{$R_{1} (\rm{R_{\odot}})$} & \multicolumn{1}{c}{$\Sigma r^{2} (I)$}  \\
\hline
$90^{\circ}$  & $475.000 \pm 3.171$  & $0.215 \pm 0.005$  & $5.005 \pm 0.022$  & $4.462 \pm 0.033$  & $0.437 \pm 0.008$  & 6.58 &  2.87 &  124.18 & $0.147 \times 10^{-14}$ \\
$80^{\circ}$  & $474.346 \pm 3.290$  & $0.215 \pm 0.005$  & $5.009 \pm 0.022$  & $4.476 \pm 0.035$  & $0.448 \pm 0.009$  & 6.50 &  2.92 &  124.06 & $0.146 \times 10^{-14}$ \\
$70^{\circ}$  & $471.506 \pm 2.823$  & $0.214 \pm 0.005$  & $5.026 \pm 0.022$  & $4.512 \pm 0.029$  & $0.485 \pm 0.008$  & 6.23 &  3.02 &  123.72 & $0.144 \times 10^{-14}$ \\
$60^{\circ}$  & $467.796 \pm 2.655$  & $0.210 \pm 0.005$  & $5.052 \pm 0.022$  & $4.594 \pm 0.027$  & $0.556 \pm 0.009$  & 5.80 &  3.23 &  123.14 & $0.142 \times 10^{-14}$ \\
$50^{\circ}$  & $464.284 \pm 2.719$  & $0.201 \pm 0.005$  & $5.077 \pm 0.023$  & $4.758 \pm 0.030$  & $0.689 \pm 0.011$  & 5.23 &  3.60 &  122.29 & $0.145 \times 10^{-14}$ \\
$40^{\circ}$  & $463.957 \pm 3.208$  & $0.170 \pm 0.005$  & $5.070 \pm 0.028$  & $5.087 \pm 0.041$  & $0.949 \pm 0.019$  & 4.52 &  4.29 &  121.09 & $0.169 \times 10^{-14}$ \\
$30^{\circ}$  & $472.820 \pm 3.927$  & $0.108 \pm 0.004$  & $4.982 \pm 0.037$  & $5.878 \pm 0.058$  & $1.579 \pm 0.036$  & 3.62 &  5.71 &  119.24 & $0.276 \times 10^{-14}$ \\
\dag \ $60^{\circ}$  & $465.730 \pm 2.482$  & $0.212 \pm 0.005$  & $5.053 \pm 0.021$  & $4.594 \pm 0.025$  & $0.560 \pm 0.008$  & 5.72 &  3.20 &  122.76 & $0.141 \times 10^{-14}$ \\
\hline
\end{tabular}
\end{table*}

\subsubsection{OGLE053124.49}

\begin{figure}
\begin{center}
\includegraphics[width=0.5\textwidth]{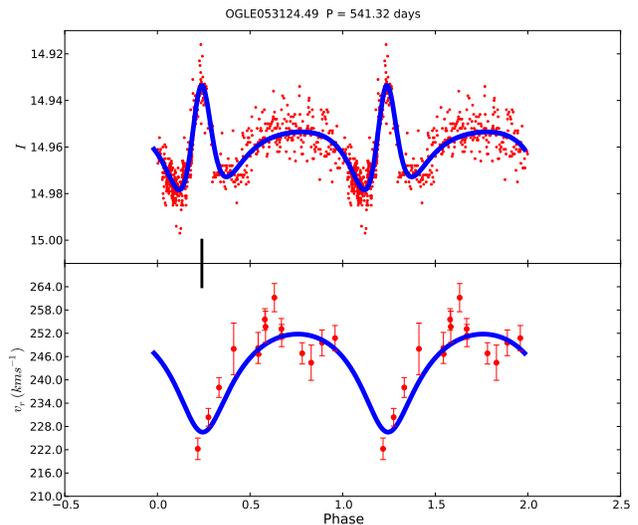}
\end{center}
\caption[]{Observed OGLE $I$ light curve and observed radial velocity curve for OGLE053124.49 (red points) and modelled light and velocity curves (blue lines) at an inclination of $70^{\circ}$. The vertical black line marks the phase of periastron.}
\label{3124soln}
\end{figure}

The observed light and velocity curves are shown in Fig.~\ref{3124soln}, from which it is clear that this star also shows unequal maxima and minima in its light curve. At superior conjunction, the relevant minimum of the light curve is deeper, i.e.\ the star is dimmer on the end nearest the companion, as expected due to gravity darkening.

The angle of periastron is $\sim 176^{\circ}$, so periastron occurs around the same time as the brighter maximum, at a phase of $\sim 0.24$. Again the narrow maximum can be attributed to the star moving fast at periastron in a highly eccentric orbit, but this does not explain why this maximum is brighter than the other.

The results of the WD modelling are shown in Table~\ref{3124table}. Semimajor axis, $e$, and $\omega$ all vary slightly with decreasing $i$, while $\Omega_1$ and $q$ vary more significantly with $i$. 

The best solution with a compact secondary is at $90^{\circ}$, according to the sum of squares of residuals of the light curve. However using the \cite{girardi} evolutionary tracks as above to find the temperature and radius of the MS companion resulted in eclipses of the light curve at $90^{\circ}$ and $80^{\circ}$. The eclipses disappeared at $i = 70^{\circ}$, where the red giant has a mass of $2.0\,M_{\odot}$, and a radius of $92\,R_{\odot}$. From the colour temperature of 3880\,K, the luminosity is $1722\,L_{\odot}$. From \cite{girardi}, a red giant of this mass and luminosity and at LMC metallicity has an age of $\sim 1.3 \times 10^{9}$ y, and is an AGB star around the time of its first thermal pulse. At the same age, the $1\,M_{\odot}$ MS companion should have $L = 1.19\,L_{\odot}$, T = 6120\,K, and $R = 0.97\,R_{\odot}$. We re-ran the code at $i = 70^{\circ}$ with $T_{\rm{eff _2}}$ increased to 6120\,K and $\Omega_2$ altered so $R_2 = 0.97\,R_{\odot}$. This solution is shown in the last row of Table~\ref{3124table} and in Fig.~\ref{3124soln}. 

\begin{table*}
\centering
\caption[]{Simultaneous light and velocity solutions for OGLE053124.49. \dag \ denotes the final solution at the best inclination and with the most accurate companion properties.}
\label{3124table}
\begin{tabular}{cccccccccc}
\hline
\multicolumn{1}{c}{$i$}  &  \multicolumn{1}{c}{$a (\rm{R_{\odot}})$}  &  \multicolumn{1}{c}{$e$}  &  \multicolumn{1}{c}{$\omega$ (radians)}  &  \multicolumn{1}{c}{$\Omega_{1}$}  &  \multicolumn{1}{c}{$q$}  &  \multicolumn{1}{c}{$M_{1} (\rm{M_{\odot}})$}  &  \multicolumn{1}{c}{$M_{2} (\rm{M_{\odot}})$}  &  \multicolumn{1}{c}{$R_{1} (\rm{R_{\odot}})$} & \multicolumn{1}{c}{$\Sigma r^{2} (I)$}  \\
\hline     
$90^{\circ}$  & $407.136 \pm 6.991$ &  $0.286 \pm 0.006$  & $3.071 \pm 0.019$  & $5.188 \pm 0.113$  & $0.467 \pm 0.029$ &  2.11 &  0.99 &  93.00  & $0.352 \times 10^{-15}$ \\
$80^{\circ}$  & $405.453 \pm 5.977$ &  $0.288 \pm 0.006$  & $3.071 \pm 0.019$  & $5.193 \pm 0.097$  & $0.476 \pm 0.025$ &  2.07 &  0.99 &  92.86  & $0.353 \times 10^{-15}$ \\
$70^{\circ}$  & $403.084 \pm 5.348$ &  $0.294 \pm 0.006$  & $3.072 \pm 0.019$  & $5.258 \pm 0.087$  & $0.518 \pm 0.023$ &  1.98 &  1.02 &  92.40  & $0.356 \times 10^{-15}$ \\
$60^{\circ}$  & $399.950 \pm 5.083$ &  $0.302 \pm 0.007$  & $3.079 \pm 0.019$  & $5.389 \pm 0.085$  & $0.595 \pm 0.025$ &  1.84 &  1.09 &  91.64  & $0.379 \times 10^{-15}$ \\
$50^{\circ}$  & $399.535 \pm 5.523$ &  $0.306 \pm 0.008$  & $3.107 \pm 0.020$  & $5.653 \pm 0.098$  & $0.733 \pm 0.032$ &  1.69 &  1.24 &  90.59  & $0.465 \times 10^{-15}$ \\
$40^{\circ}$  & $408.102 \pm 7.726$ &  $0.295 \pm 0.012$  & $3.194 \pm 0.027$  & $6.144 \pm 0.153$  & $0.972 \pm 0.059$ &  1.58 &  1.54 &  89.37  & $0.699 \times 10^{-15}$ \\
$30^{\circ}$  & $377.018 \pm 8.088$ &  $0.083 \pm 0.014$  & $4.184 \pm 0.087$  & $6.516 \pm 0.161$  & $1.700 \pm 0.086$ &  0.91 &  1.55 &  86.47  & $1.150 \times 10^{-15}$ \\
\dag \ $70^{\circ}$  & $402.608 \pm 6.848$ &  $0.292 \pm 0.006$  & $3.076 \pm 0.018$  & $5.254 \pm 0.116$  & $0.519 \pm 0.032$ &  1.97 &  1.02 &  92.40  & $0.356 \times 10^{-15}$ \\
\hline
\end{tabular}
\end{table*}

\subsubsection{OGLE053159.96}

\begin{figure}
\begin{center}
\includegraphics[width=0.5\textwidth]{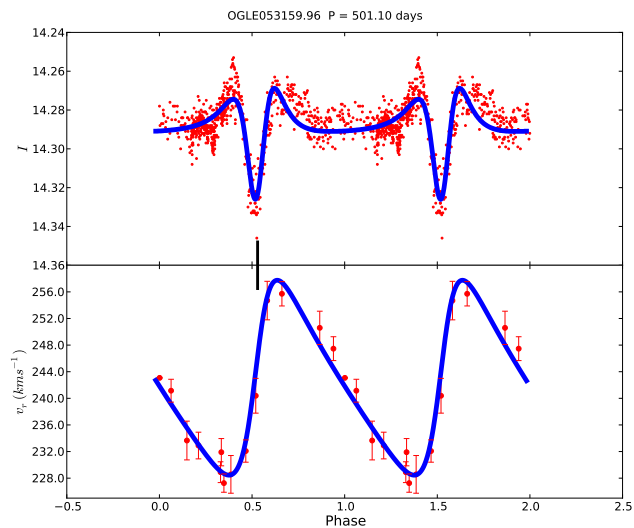}
\end{center}
\caption[]{Observed OGLE $I$ light curve and observed radial velocity curve for OGLE053159.96 (red points) and modelled light and velocity curves (blue lines) at an inclination of $60^{\circ}$. The vertical black line marks the phase of periastron.}
\label{3159soln}
\end{figure}

The observed light and velocity curves are shown in Fig.~\ref{3159soln}. This star has a light curve with unequal maxima and minima, and again its deeper light minimum occurs at inferior conjunction.

The angle of periastron is $\sim 280^{\circ}$, so periastron occurs between inferior conjunction and the subsequent light maximum, at a phase of $\sim 0.53$. In this case, periastron has squeezed the deeper minimum in phase, rather than one of the maxima.

The solutions are shown in Table~\ref{3159table}. Semimajor axis, $e$, and $\omega$ all vary slowly with $i$. Primary potential and $q$ vary significantly with $i$. The stepping down of $i$ for successive solutions was halted before $q$ exceeded unity, as at $i=40^{\circ}$ \textsc{dc} was unable to fit the full amplitude of the light curve.

According to the sum of squares of residuals of the light curve, the best solution is at $i=80^{\circ}$. However using a more accurate temperature and radius for the MS companion results in eclipses at $80^{\circ}$ and $70^{\circ}$, due to the increased radius of the MS star. The eclipses disappeared at $i = 60^{\circ}$, where the red giant has a mass of $4.8\,M_{\odot}$, and a radius of $125\,R_{\odot}$. With a colour temperature of 3880\,K, the red giant's luminosity is $3179\,L_{\odot}$. Using the data of \cite{girardi}, a red giant of this mass and luminosity and LMC metallicity should have an age of $\sim 1.05 \times 10^{8}$ y, and be at or near the RGB tip. At the same age, the $2.1\,M_{\odot}$ MS companion should have $L = 35\,L_{\odot}$, T = 10590\,K, and $R = 1.76\,R_{\odot}$. We further refined our best solution ($i = 60^{\circ}$) with $T_{\rm{eff _2}}$ increased to 10590\,K and $\Omega_2$ altered so $R_2 = 1.76\,R_{\odot}$. This solution is shown in the last row of Table~\ref{3159table} and in Fig.~\ref{3159soln}. This star also has higher mass than expected for the general LMC field population.

\begin{table*}
\centering
\caption[]{Simultaneous light and velocity solutions for OGLE053159.96. \dag \ denotes the final solution at the best inclination and with the most accurate companion properties.}
\label{3159table}
\begin{tabular}{cccccccccc}
\hline
\multicolumn{1}{c}{$i$}  &  \multicolumn{1}{c}{$a (\rm{R_{\odot}})$}  &  \multicolumn{1}{c}{$e$}  &  \multicolumn{1}{c}{$\omega$ (radians)}  &  \multicolumn{1}{c}{$\Omega_{1}$}  &  \multicolumn{1}{c}{$q$}  &  \multicolumn{1}{c}{$M_{1} (\rm{M_{\odot}})$}  &  \multicolumn{1}{c}{$M_{2} (\rm{M_{\odot}})$}  &  \multicolumn{1}{c}{$R_{1} (\rm{R_{\odot}})$} & \multicolumn{1}{c}{$\Sigma r^{2} (I)$}  \\
\hline     
$90^{\circ}$  & $525.077 \pm 4.739$ &  $0.405 \pm 0.008$  & $4.874 \pm 0.012$  & $4.859 \pm 0.044$  & $0.336 \pm 0.088$  & 5.80  & 1.95  & 126.84  & $0.19 \times 10^{-14}$ \\
$80^{\circ}$  & $522.302 \pm 7.409$ &  $0.405 \pm 0.008$  & $4.866 \pm 0.012$  & $4.856 \pm 0.081$  & $0.342 \pm 0.016$  & 5.68  & 1.94  & 126.65  & $0.19 \times 10^{-14}$ \\
$70^{\circ}$  & $516.514 \pm 5.895$ &  $0.400 \pm 0.007$  & $4.872 \pm 0.012$  & $4.886 \pm 0.062$  & $0.375 \pm 0.013$  & 5.36  & 2.01  & 126.05  & $0.19 \times 10^{-14}$ \\
$60^{\circ}$  & $505.586 \pm 5.025$ &  $0.390 \pm 0.007$  & $4.881 \pm 0.013$  & $4.937 \pm 0.052$  & $0.436 \pm 0.013$  & 4.81  & 2.10  & 125.03  & $0.19 \times 10^{-14}$ \\
$50^{\circ}$  & $492.890 \pm 4.874$ &  $0.367 \pm 0.007$  & $4.898 \pm 0.016$  & $5.062 \pm 0.050$  & $0.550 \pm 0.015$  & 4.13  & 2.27  & 123.48  & $0.21 \times 10^{-14}$ \\
$40^{\circ}$  & $489.737 \pm 7.847$ &  $0.305 \pm 0.009$  & $4.946 \pm 0.028$  & $5.380 \pm 0.103$  & $0.778 \pm 0.038$  & 3.53  & 2.75  & 121.42  & $0.27 \times 10^{-14}$ \\
\dag \ $60^{\circ}$  & $505.150 \pm 6.698$ &  $0.391 \pm 0.007$  & $4.877 \pm 0.013$  & $4.948 \pm 0.079$  & $0.439 \pm 0.020$  & 4.79  & 2.10  & 124.72  & $0.19 \times 10^{-14}$ \\
\hline
\end{tabular}
\end{table*}

\section{Discussion}

All the stars in our sample clearly show the doubling of the light curve with respect to the velocity curve that is the hallmark of ellipsoidal variation. All the light curves display minima of unequal depths, another common property of ellipsoidal variables. In variables with circular orbits, the deeper minimum is caused by gravity darkening on the inner end of the ellipsoidal red giant. In that case the deeper minimum should be found at superior conjunction, however for most of our eccentric sample the opposite is true. 

The majority of stars in our sample also show maxima of unequal heights in their light curves, a phenomenon that was not seen in ellipsoidal variables with small or zero eccentricity \citep{seqEpaper}. A possible explanation for the unequal maxima and the unexpected placement of the deeper minimum in eccentric ellipsoidal variables is presented in Section~\ref{distortion}.

Many stars in our sample are more massive and more luminous than the average LMC red giant. According to the evolutionary tracks of \cite{girardi}, two are AGB stars. This is likely to be mostly due to a luminosity selection effect, as we selected the brightest eccentric candidates from the OGLE database to obtain targets suitable for our observing facilities. Selecting variables with high eccentricity is also likely to mean longer average orbital periods and hence higher luminosities, since \nocite{ogleellipsoidal} Soszy\'nski et al. (2004b) noted that the eccentric ellipsoidal variables in their sample generally had longer periods than the low-eccentricity variables. We note that it is unclear why higher eccentricity and longer periods should be linked in the case of ellipsoidal variables, since tidal circularisation time depends not on orbital separation (i.e. period) but very sensitively on fractional lobe filling ($a/R$). 

\begin{figure}
\begin{center}
\includegraphics[width=0.5\textwidth]{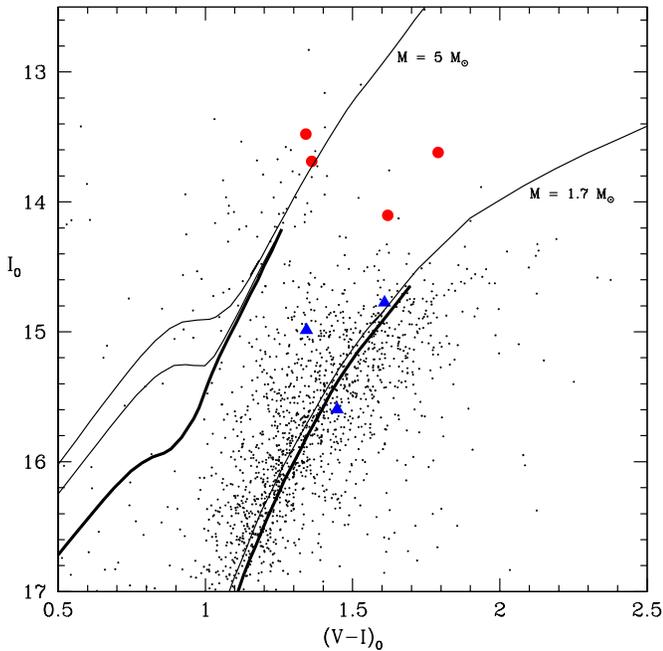}
\end{center}
\caption[]{The sequence E variables from \nocite{ogleellipsoidal} Soszy\'nski et al. (2004b) in the ($I_0,(V-I)_0$) plane (small dots). The stars studied in this paper are shown as blue triangles when $M < 2\,M_{\odot}$ and as large red dots when $M > 4\,M_{\odot}$.  Also shown are evolutionary tracks from \cite{bertelli08,bertelli09}. The RGB (i.e. up to He core ignition) is shown by thick lines. The distance modulus and reddening are as described in Section~\ref{model}.  The \citeauthor{bertelli09} tracks have been shifted 0.1 mag. bluer in $V-I$ to match the observed $V-I$.}
\label{hrdiag}
\end{figure}

Fig.~\ref{hrdiag} shows sequence E variables in the OGLE II database in the ($I_0,(V-I)_0$) plane. The majority of the sequence E stars are on the low mass RGB, but there are significant numbers of more massive stars around $(V-I,I)=(1.2,14)$, where the more massive members of our sample lie. These may be useful for future studies of intermediate mass stars on the early AGB.

The inclinations of our sample are between $50^{\circ}$ and $90^{\circ}$, with a mean inclination of $66^{\circ}$. A bias towards high inclinations is expected for ellipsoidal variation, which should not be detectable at very low inclinations, although ellipsoids with a large fractional lobe filling may be visible as low-amplitude variables at inclinations as low as $30^{\circ}$. It is interesting to note that although ellipsoidal variation would certainly be visible in edge-on orbits, in our sample of variables without eclipses, inclinations this high are unlikely unless the companion is particularly small. However in many of our systems, both components are intermediate mass stars ($M \geq 1.85\,M_{\odot}$), meaning that the radius ratio of red giant to main sequence star is not very large (as it is for low mass stars, $M \leq 1.85\,M_{\odot}$). In this situation, the companions will cause easily detected eclipses, so by selecting non-eclipsing variables we have unwittingly selected fewer edge-on orbits. In a more representative sample of non-eclipsing ellipsoidal variables with lower average mass (closer to the LMC average), and with relatively smaller companion radii that do not cause observable eclipses even in edge-on orbits, we suspect the mean orbital inclination would be higher.

\begin{figure}
\begin{center}
\includegraphics[width=0.5\textwidth]{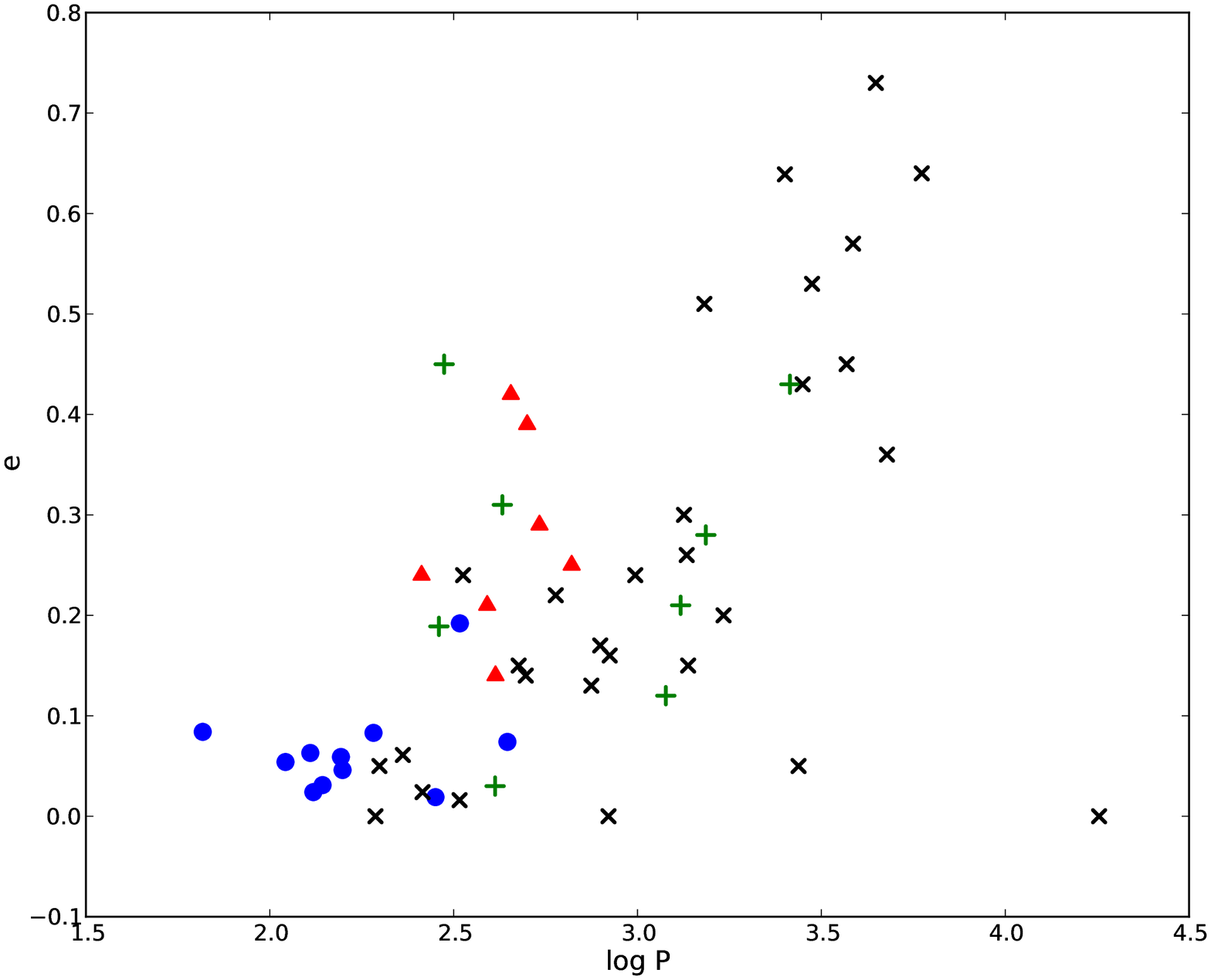}
\end{center}
\caption[]{Eccentricity vs. period for the current sample of eccentric ellipsoidal variables (red triangles); the low-$e$ ellipsoidal sample of \cite{seqEpaper} (blue circles); post-AGB stars from \cite{waters93,vanwinckel95,vanwinckel98,vanwinckel99,pollardcottrell}; and \cite{gonzalezwallerstein} (green pluses); and M giant binaries from \cite{famaey09} (black crosses).}
\label{evsp}
\end{figure}

We have confirmed the hypothesis of \nocite{ogleellipsoidal} Soszy\'nski et al. (2004b), that ellipsoidal light curves with strange shapes represent eccentric orbits. The eccentricities of all stars in our sample are significantly nonzero and generally high, ranging from 0.14 to 0.42. The mean eccentricity of the sample is 0.28. Fig.~\ref{evsp} shows the location of these eccentric ellipsoidal variables in the $(e, \log P)$ plane compared to other evolved binaries. At short periods, eccentricities are lower as expected from tidal theory. This graph is a vivid demonstration of the surprisingly high eccentricities found amongst many evolved binaries. According to \cite{rob10}, population synthesis studies of Ba stars predict all orbits $< 4000$ days should be circular, a result that is at complete odds with observations. The diverse range of periods, masses and evolutionary states of binaries that seem to have somehow escaped the circularising effect of tidal forces serves to reinforce the need for an understanding of the mechanism that maintains or increases eccentricity.

\begin{figure}
\begin{center}
\includegraphics[width=0.5\textwidth]{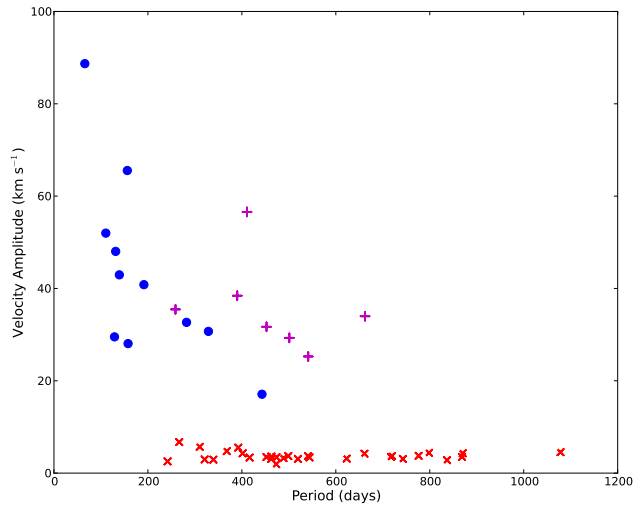}
\end{center}
\caption[]{Full velocity amplitude vs. period for the current sample of eccentric ellipsoidal variables (magenta pluses), the mostly circular ellipsoidal sample of \cite{seqEpaper} (blue circles), and the LSPV sample of \cite{seqDpaper} (red crosses).}
\label{vfampvsp}
\end{figure}

The full velocity amplitudes of our sample lie between 25 and 56 $\rm{km\,s^{-1}}$, with a mean of 35.8 $\rm{km\,s^{-1}}$. Fig.~\ref{vfampvsp} shows that the current sample of ellipsoidal variables with high eccentricities falls within the velocity amplitude distribution of the ellipsoidal variables with mostly circular orbits presented in \cite{seqEpaper}. However the eccentric variables generally lie at longer periods. Also shown in Fig.~\ref{vfampvsp} are a sample of LSPVs from \cite{seqDpaper}, once again demonstrating the marked difference between stars with Long Secondary Periods and ellipsoidal variables.

Two individual stars are worthy of comment. OGLE052850.12 is the brightest and coolest star in our sample, with the largest light amplitude, meaning it is nearly filling its Roche Lobe. It is of relatively high mass ($\sim 4.2\,M_{\odot}$) but not the highest in our sample. According to the evolutionary tracks of \cite{girardi} it is an AGB star, which explains the obvious pulsations in its light curve. There is a very real possibility that this object will fill its Roche Lobe before it reaches the AGB tip and become a PN, possibly of asymmetric shape, with a close binary central star.

OGLE053124.49 is the most evolved star in our sample, and according to the evolutionary tracks is an AGB star near the time of its first thermal pulse. It has one of the smallest light amplitudes in our sample, suggesting a low fractional lobe filling (as its inclination is similar to that of OGLE052850.12). Given its advanced evolutionary state and low fractional lobe filling, it is possible that this star may not fill its Roche Lobe before it reaches the AGB tip and may lose its envelope via the superwind as single AGB stars do, leaving a remnant PN with a wide binary companion.

\subsection{Pulsation of the Ellipsoidal Red Giant}

The two AGB stars in our sample show brief intervals of pulsation in their light curves. OGLE052850.12 has two separate pulsation episodes at a phase near 0.1 when the pulsation period is 29.5 days, and a phase near 0.8 when the pulsation period is 26.4 days.  These periods were determined using the task \textit{pdm} in \textsc{iraf}.  OGLE053124.49 shows evidence for pulsation at one interval near phase 0.3 when the period is 19.0 days.

Because $L$, $T_{\rm{eff}}$ and $M$ have all been reliably determined for these red giants, they provide a unique opportunity to find the modes of oscillation involved, assuming these modes are radial. As mentioned above, OGLE052850.12 is a star on the early AGB with essentially all its luminosity coming from the helium burning shell \cite{bertelli09}.  It is currently undergoing second dredge-up. Similarly, the evolutionary tracks of \cite{bertelli08} confirm that the $1.97\,M_{\odot}$ star OGLE053124.49 does not develop a degenerate core on the first ascent of the giant branch and only reaches the observed luminosity of $1722\,L_{\odot}$ when in the thermally pulsing AGB stage.

The linear pulsation code described in \cite{foxwood82} with updated opacities was used to calculate the periods of the first four modes of radial pulsation in these stars, and they are given in Table~\ref{pulstab}.  It is clear that OGLE052850.12 is pulsating in the second overtone while OGLE053124.49 could be pulsating in the first or second overtone.

\begin{table*}
\centering
\caption{Periods of the first four radial pulsation modes for stars with evidence of pulsation.}
\label{pulstab}
\begin{tabular}{lcccccccc}
\hline
\multicolumn{1}{c}{Star}  & \multicolumn{1}{c}{$P$ (days)}  & \multicolumn{1}{c}{$M\ (M_{\odot})$}  & \multicolumn{1}{c}{$L\ (L_{\odot})$}  & \multicolumn{1}{c}{$T_{\rm{eff}}$}  & \multicolumn{1}{c}{$P_0$}  & \multicolumn{1}{c}{$P_1$}  & \multicolumn{1}{c}{$P_2$}  & \multicolumn{1}{c}{$P_3$} \\	
\hline
OGLE052850.12 & 29.5,26.4 & 4.21 & 5325 & 3720 & 74.3 & 41.5 & 29.7 & 22.0 \\
OGLE053124.49 & 19.0         & 1.97 & 1722 & 3880 & 36.8 & 22.6 & 16.2 & 12.5 \\
\hline
\end{tabular}
\end{table*}

\begin{figure}
\begin{center}
\includegraphics[width=0.5\textwidth]{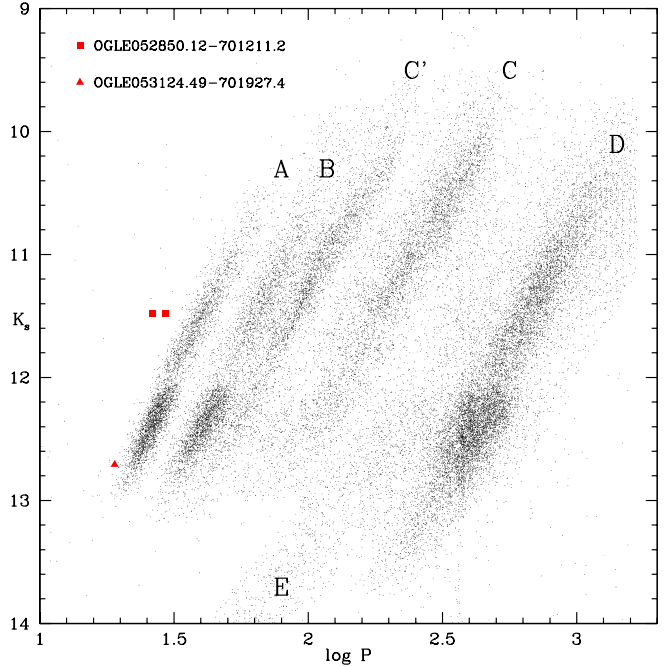}
\end{center}
\caption[]{The position of OGLE052850.12 and OGLE053124.49 in the $K$--$\log P$ diagram for variable red giant stars in the LMC, using data from \cite{fraser08}. The sequences are labelled according to the scheme of \cite{ita04}.}
\label{k-logp}
\end{figure}

The two stars are shown in the $K$--$\log P$ diagram in Fig.~\ref{k-logp} along with the population of variable red giants in the LMC from \cite{fraser08}.  In this figure, sequence C
corresponds to the fundamental mode of radial pulsation and sequences C$'$, B and A correspond to successively higher order modes \nocite{wood99mn} (Wood et al.\ 1999).  Pulsation models predict unambiguously that sequence C$'$ is the first overtone but sequences B and A may be the third and fifth overtone, respectively, rather than the second and third overtone \citep{woodarnett}.  In any case, one would expect OGLE052850.12 and OGLE053124.49 to lie between sequences C and C$'$ since these two stars pulsate in the second overtone, or possibly the first overtone in the case of OGLE053124.49.  They clearly do not lie in this position. In fact, their periods indicate that if they are similar to typical LMC red giants, they should be pulsating in high overtones.

The reason for such short periods in these stars is that their masses are larger than the typical field stars in the LMC which make up the bulk of the stars in Fig.~\ref{k-logp}.  For overtone periods, the period varies to good approximation as $P \propto R^{1.5} M^{-0.5}$ \citep[e.g.][]{foxwood82}.  If this is combined with the variation of $T_{\rm{eff}}$ with mass given by equation 6 in \cite{wood90} and the definition of effective temperature $L = 4 \pi \sigma R^2T_{\rm{eff}}^4$, we find that $P \propto M^{-1.01}$.  Assuming the LMC field population has a mean mass of $1.5\,M_{\odot}$ \citep[e.g.][]{bertelli92}, the 29 day second overtone pulsation of the $4.21\,M_{\odot}$ star OGLE052850.12 would become an 82.2 day ($\log P=1.91$) overtone pulsation in a $1.5\,M_{\odot}$ LMC field star. This period lies between sequences C$'$ and B and would correspond to the second overtone, according to \cite{woodarnett}.  Thus for this star, the large mass can explain the position in the $K$--$\log P$ diagram relative to other LMC red giants.

For OGLE053124.49, the mass of $1.97\,M_{\odot}$ and the arguments above would only move the star onto sequence A if it were a $1.5\,M_{\odot}$ field star, corresponding to the third to fifth overtone. To shift to the first or second overtone would require the LMC field population at the luminosity of OGLE053124.49 to have a mass of $\sim 1\,M_{\odot}$.  This suggests that as the luminosity increases along each of the period--luminosity sequences in Fig.~\ref{k-logp}, the mass increases.  This is consistent with theoretical models which show that higher mass tends to stabilise red giant pulsation and hence higher luminosities are required to make higher mass red giants unstable.  It is also consistent with studies which show that Mira variables of longer periods have higher masses \citep[e.g.][]{feast63,woodsebo}.

The $K$--$\log P$ relations for LMC red giants by \cite{oglep-l} show a weak sequence of stars on the short period side of sequence A, just where OGLE052850.12 and OGLE053124.49 lie.  Our results show that this sequence could be made up of intermediate mass stars that do not ascend the RGB.  These stars would be pulsating in the first or second overtone rather than in an overtone higher than that corresponding to sequence A.

\subsection{Increased Distortion in High Eccentricity Orbits}
\label{distortion}

It has been noted above that many of the eccentric ellipsoidal variables in our sample show unequal maxima as well as unequal minima in their light curves. As \cite{seqEpaper} showed, in normal circumstances the deeper minimum is caused by gravity darkening of the red giant on the side nearest its companion and should therefore occur at superior conjunction. In this regime there is also no reason why the light maxima should be unequal.

From a glance at the phased light and velocity curves of the current sample, it is clear that for many stars, the deeper light minimum occurs at inferior, not superior, conjunction. This suggests that when the `outer end' of the ellipsoidal red giant is facing us, the star appears dimmer due to some effect outweighing the gravity darkening of the `inner end'. 

A related phenomenon occurs with maxima. For 6 of the 7 stars, one of the light maxima is narrower and brighter than the other. The narrowness of this maximum -- its shorter span in time -- can be explained by its proximity to periastron and the eccentricity of the orbit. At periastron in a highly eccentric orbit, the star moves significantly faster than at apastron, so that part of the light curve appears `squashed' in phase. But what causes a maximum to be brighter as well as narrower when it lies close to periastron?

The results of our modelling show that the deeper light minimum and the brighter maximum are always close to periastron. Given the high eccentricity of these orbits, here we suggest that the cause of these unequal maxima and minima, and the effect that can outweigh gravity darkening of the inner end of the ellipsoid, is increased distortion of the red giant at periastron. 

Due to small periastron separations resulting from highly eccentric orbits, at periastron the greater influence of the companion causes the red giant to become more distorted, its ellipsoidal shape more pronounced. As the major axis of the ellipsoid lengthens and the minor axes contract, the apparent surface area of the ellipsoid seen `side-on' increases, and its apparent surface area seen `end-on' decreases. Since the light maxima of ellipsoidal variables correspond to observing the star side-on, the increased apparent surface area means higher observed flux, so the maximum closest to periastron is brighter. Similarly, light minima are seen when the ellipsoid is end-on, and a decreased apparent surface area means lower observed flux, so the minimum nearest periastron is dimmer than the alternate minimum.

This distortion effect would naturally become more pronounced with eccentricity, and be less apparent for low eccentricity orbits and non-existent for circular orbits. It would also depend on the fractional lobe filling of the red giant, and where in the orbit periastron occurs, i.e.\ the angle of periastron. For example greater fractional lobe filling could produce high distortion in an only moderately eccentric system, and higher distortion could also be observed if periastron was coincident with a light maximum or minimum. This complex dependency is hinted at in the light curves of OGLE052013.51 and OGLE052850.12 (Figs.~\ref{2013soln} and \ref{2850soln} respectively). Although OGLE052850.12 has a lower eccentricity than OGLE052013.51, their light curves show a similar magnitude difference between the brighter and dimmer maxima, due to the fact that OGLE052850.12 has a greater fractional lobe filling. Further, although OGLE052438.40 has the lowest eccentricity of the current sample, periastron is almost coincident with the inferior conjunction light minimum, causing the distortion effect to result in a deeper minimum at inferior conjunction than at superior. However, for the nearly circular orbits of the ellipsoidal variables in the \cite{seqEpaper} sample, the maxima are equal and the deeper minimum is purely due to gravity darkening and is found, as expected, at superior conjunction.

We used the WD code to quantitatively evaluate our hypothesis that the unequal maxima and asymmetric shapes of these light curves are caused by apparent surface area variations resulting from increased distortion at periastron. Using each star's calculated final solution parameters and running \textsc{lc} in image mode allowed us to obtain the apparent surface area at each calculated orbital phase point. In the absence of any variation of $T_{\rm{eff}}$ or limb darkening, we would expect the apparent surface area variation to closely mimic the light curve. As expected, the apparent surface area variation was generally remarkably similar to the light variation, with only slight differences in phase and amplitude. We assume these differences are due to the differing contribution of limb darkening as the star rotates throughout its orbit. Therefore it seems that increased distortion at periastron is indeed responsible for the variety of asymmetric light curve shapes observed in eccentric ellipsoidal variables.

\subsection{Maintaining Eccentricity in Evolved Close Binaries}

The high eccentricities of our current sample serve to reiterate the fact that current tidal theory cannot accurately explain orbital evolution. The tidal circularisation time for our close red giant binaries is orders of magnitude smaller than the average lifetime of the ellipsoidal phase, yet eccentric orbits are not uncommon. This suggests that orbital eccentricity is maintained or increased in evolved binaries by some unknown mechanism. 

As we noted in Section~\ref{intro}, two of the proposed mechanisms are mass transfer at periastron \citep{soker00} and interaction with a circumbinary disk \citep{arty}. Both of these may be tested observationally by searching for the signatures of circumstellar matter or accretion. We showed in \cite{seqEpaper} that sequence E binaries show no evidence of a mid-infrared excess that would indicate enhanced circumstellar or circumbinary dust. Thus, there is no observed evidence for mass loss or disks in these binary systems. However, mass transfer between the components cannot be ruled out. 

By modelling eccentric ellipsoidal red giant binaries, we have determined complete orbital solutions for these poorly understood stars. Our results can serve as input for future hydrodynamic modelling to determine how eccentricity is maintained in these stars, and in other evolved eccentric binaries, possibly by mass transfer.

\section{Conclusions}

We have confirmed that ellipsoidal red giant binaries with unusual light curve shapes are in eccentric orbits, and we have used the Wilson--Devinney code to model the orbits and obtain orbital and stellar parameters, including masses of the stars. We find ellipsoidal variables that do not display eclipses are generally at high orbital inclinations, although edge-on orbits amongst intermediate-mass non-eclipsing ellipsoidal variables are rare. Unlike their counterparts in circular orbits, eccentric ellipsoidal variables generally have unequal maxima as well as minima in their light curves, often with one maximum spanning a significantly narrower phase and the deeper minimum occurring at inferior conjunction, instead of at superior conjunction as gravity darkening would dictate. We inferred that these phenomena are due to greater distortion of the ellipsoidal red giant at periastron due to the high eccentricities, a hypothesis that is supported by the modelled apparent surface area. By determining the properties of eccentric sequence E stars we have laid the groundwork for future hydrodynamic modelling to determine how the eccentricity is maintained in these stars. Finally, we showed that the pulsation found in two of the red giants corresponds to the first or second overtone.  In the $K$--$\log P$ diagram for pulsating red giants, these stars have periods shorter than sequence A because their masses are higher than those of the typical LMC field population.

\bibliographystyle{mn2e}
\bibliography{bibliographynew}

\begin{thebibliography}{}

\bibitem[\protect\citeauthoryear{{Artymowicz}, {Clarke}, {Lubow} \&
  {Pringle}}{{Artymowicz} et~al.}{1991}]{arty}
{Artymowicz} P.,  {Clarke} C.~J.,  {Lubow} S.~H.,    {Pringle} J.~E.,  1991,
  \apjl, 370, L35

\bibitem[\protect\citeauthoryear{{Bertelli}, {Girardi}, {Marigo} \&
  {Nasi}}{{Bertelli} et~al.}{2008}]{bertelli08}
{Bertelli} G.,  {Girardi} L.,  {Marigo} P.,    {Nasi} E.,  2008, \aap, 484, 815

\bibitem[\protect\citeauthoryear{{Bertelli}, {Mateo}, {Chiosi} \&
  {Bressan}}{{Bertelli} et~al.}{1992}]{bertelli92}
{Bertelli} G.,  {Mateo} M.,  {Chiosi} C.,    {Bressan} A.,  1992, \apj, 388,
  400

\bibitem[\protect\citeauthoryear{{Bertelli}, {Nasi}, {Girardi} \&
  {Marigo}}{{Bertelli} et~al.}{2009}]{bertelli09}
{Bertelli} G.,  {Nasi} E.,  {Girardi} L.,    {Marigo} P.,  2009, \aap, 508, 355

\bibitem[\protect\citeauthoryear{{Bessell}}{{Bessell}}{1979}]{bessell79}
{Bessell} M.~S.,  1979, \pasp, 91, 589

\bibitem[\protect\citeauthoryear{{Cardelli}, {Clayton} \& {Mathis}}{{Cardelli}
  et~al.}{1989}]{cardelli}
{Cardelli} J.~A.,  {Clayton} G.~C.,    {Mathis} J.~S.,  1989, \apj, 345, 245

\bibitem[\protect\citeauthoryear{{Famaey}, {Pourbaix}, {Frankowski}, {van Eck},
  {Mayor}, {Udry} \& {Jorissen}}{{Famaey} et~al.}{2009}]{famaey09}
{Famaey} B.,  {Pourbaix} D.,  {Frankowski} A.,  {van Eck} S.,  {Mayor} M.,
  {Udry} S.,    {Jorissen} A.,  2009, \aap, 498, 627

\bibitem[\protect\citeauthoryear{{Feast}}{{Feast}}{1963}]{feast63}
{Feast} M.~W.,  1963, \mnras, 125, 367

\bibitem[\protect\citeauthoryear{{Fox} \& {Wood}}{{Fox} \&
  {Wood}}{1982}]{foxwood82}
{Fox} M.~W.,  {Wood} P.~R.,  1982, \apj, 259, 198

\bibitem[\protect\citeauthoryear{{Fraser}, {Hawley} \& {Cook}}{{Fraser}
  et~al.}{2008}]{fraser08}
{Fraser} O.~J.,  {Hawley} S.~L.,    {Cook} K.~H.,  2008, \aj, 136, 1242

\bibitem[\protect\citeauthoryear{{Fraser}, {Hawley}, {Cook} \&
  {Keller}}{{Fraser} et~al.}{2005}]{fraser05}
{Fraser} O.~J.,  {Hawley} S.~L.,  {Cook} K.~H.,    {Keller} S.~C.,  2005, \aj,
  129, 768

\bibitem[\protect\citeauthoryear{{Girardi}, {Bressan}, {Bertelli} \&
  {Chiosi}}{{Girardi} et~al.}{2000}]{girardi}
{Girardi} L.,  {Bressan} A.,  {Bertelli} G.,    {Chiosi} C.,  2000, \aaps, 141,
  371

\bibitem[\protect\citeauthoryear{{Gonzalez} \& {Wallerstein}}{{Gonzalez} \&
  {Wallerstein}}{1996}]{gonzalezwallerstein}
{Gonzalez} G.,  {Wallerstein} G.,  1996, \mnras, 280, 515

\bibitem[\protect\citeauthoryear{{Hinkle}, {Lebzelter}, {Joyce} \&
  {Fekel}}{{Hinkle} et~al.}{2002}]{hinkle02}
{Hinkle} K.~H.,  {Lebzelter} T.,  {Joyce} R.~R.,    {Fekel} F.~C.,  2002, \aj,
  123, 1002

\bibitem[\protect\citeauthoryear{{Houdashelt}, {Bell}, {Sweigart} \&
  {Wing}}{{Houdashelt} et~al.}{2000}]{houdashelt}
{Houdashelt} M.~L.,  {Bell} R.~A.,  {Sweigart} A.~V.,    {Wing} R.~F.,  2000,
  \aj, 119, 1424

\bibitem[\protect\citeauthoryear{{Ita}, {Tanab{\'e}}, {Matsunaga}, {Nakajima},
  {Nagashima}, {Nagayama}, {Kato}, {Kurita}, {Nagata}, {Sato}, {Tamura},
  {Nakaya} \& {Nakada}}{{Ita} et~al.}{2004}]{ita04}
{Ita} Y.,  {Tanab{\'e}} T.,  {Matsunaga} N.,  {Nakajima} Y.,  {Nagashima} C.,
  {Nagayama} T.,  {Kato} D.,  {Kurita} M.,  {Nagata} T.,  {Sato} S.,  {Tamura}
  M.,  {Nakaya} H.,    {Nakada} Y.,  2004, \mnras, 347, 720

\bibitem[\protect\citeauthoryear{{Izzard}, {Dermine} \& {Church}}{{Izzard}
  et~al.}{2010}]{rob10}
{Izzard} R.~G.,  {Dermine} T.,    {Church} R.~P.,  2010, \aap, 523, A10+

\bibitem[\protect\citeauthoryear{{Johnson}}{{Johnson}}{1966}]{johnson66}
{Johnson} H.~L.,  1966, \araa, 4, 193

\bibitem[\protect\citeauthoryear{{Keller} \& {Wood}}{{Keller} \&
  {Wood}}{2006}]{kellerwood}
{Keller} S.~C.,  {Wood} P.~R.,  2006, \apj, 642, 834

\bibitem[\protect\citeauthoryear{{Kurucz}}{{Kurucz}}{1993}]{kurucz}
{Kurucz} R.~L.,  1993, in {IAU Commission on Close Binary Stars} Vol.~21, {New
  atmospheres for modelling binaries and disks.}.
pp 93--101

\bibitem[\protect\citeauthoryear{{Nicholls} \& {Wood}}{{Nicholls} \&
  {Wood}}{2011}]{apn5paper}
{Nicholls} C.~P.,  {Wood} P.~R.,  2011, in Asymmetric Planetary Nebulae 5
  Conference {The connection between sequence D and sequence E red giant
  variables and asymmetric PNe}

\bibitem[\protect\citeauthoryear{{Nicholls}, {Wood} \& {Cioni}}{{Nicholls}
  et~al.}{2010}]{seqEpaper}
{Nicholls} C.~P.,  {Wood} P.~R.,    {Cioni} M.,  2010, \mnras, 405, 1770

\bibitem[\protect\citeauthoryear{{Nicholls}, {Wood}, {Cioni} \&
  {Soszy{\'n}ski}}{{Nicholls} et~al.}{2009}]{seqDpaper}
{Nicholls} C.~P.,  {Wood} P.~R.,  {Cioni} M.,    {Soszy{\'n}ski} I.,  2009,
  \mnras, 399, 2063

\bibitem[\protect\citeauthoryear{{Nie}, {Wood} \& {Nicholls}}{{Nie}
  et~al.}{2011}]{nie11}
{Nie} J.~D.,  {Wood} P.~R.,    {Nicholls} C.~P.,  2011, in preparation

\bibitem[\protect\citeauthoryear{{Pollard} \& {Cottrell}}{{Pollard} \&
  {Cottrell}}{1995}]{pollardcottrell}
{Pollard} K.~H.,  {Cottrell} P.~L.,  1995, in {R.~S.~Stobie \& P.~A.~Whitelock}
  ed., IAU Colloq. 155: Astrophysical Applications of Stellar Pulsation Vol.~83
  of Astronomical Society of the Pacific Conference Series, {The Long-term
  Variation in the RV Tauri Star U MON}.
pp 409--+

\bibitem[\protect\citeauthoryear{{Rodgers}, {Conroy} \& {Bloxham}}{{Rodgers}
  et~al.}{1988}]{2.3m}
{Rodgers} A.~W.,  {Conroy} P.,    {Bloxham} G.,  1988, \pasp, 100, 626

\bibitem[\protect\citeauthoryear{{Soker}}{{Soker}}{2000}]{soker00}
{Soker} N.,  2000, \aap, 357, 557

\bibitem[\protect\citeauthoryear{{Soszy\'nski}, {Dziembowski}, {Udalski},
  {Kubiak}, {Szymanski}, {Pietrzynski}, {Wyrzykowski}, {Szewczyk} \&
  {Ulaczyk}}{{Soszy\'nski} et~al.}{2007}]{oglep-l}
{Soszy\'nski} I.,  {Dziembowski} W.~A.,  {Udalski} A.,  {Kubiak} M.,
  {Szymanski} M.~K.,  {Pietrzynski} G.,  {Wyrzykowski} L.,  {Szewczyk} O.,
  {Ulaczyk} K.,  2007, Acta Astronomica, 57, 201

\bibitem[\protect\citeauthoryear{{Soszy{\'n}ski}, {Udalski}, {Kubiak},
  {Szymanski}, {Pietrzynski}, {Zebrun}, {Szewczyk} \&
  {Wyrzykowski}}{{Soszy{\'n}ski} et~al.}{2004}]{ogle04}
{Soszy{\'n}ski} I.,  {Udalski} A.,  {Kubiak} M.,  {Szymanski} M.,
  {Pietrzynski} G.,  {Zebrun} K.,  {Szewczyk} O.,    {Wyrzykowski} L.,  2004,
  Acta Astronomica, 54, 129

\bibitem[\protect\citeauthoryear{{Soszy{\'n}ski}, {Udalski}, {Kubiak},
  {Szymanski}, {Pietrzynski}, {Zebrun}, {Szewczyk}, {Wyrzykowski} \&
  {Dziembowski}}{{Soszy{\'n}ski} et~al.}{2004}]{ogleellipsoidal}
{Soszy{\'n}ski} I.,  {Udalski} A.,  {Kubiak} M.,  {Szymanski} M.~K.,
  {Pietrzynski} G.,  {Zebrun} K.,  {Szewczyk} O.,  {Wyrzykowski} L.,
  {Dziembowski} W.~A.,  2004, Acta Astronomica, 54, 347

\bibitem[\protect\citeauthoryear{{van Hamme}}{{van Hamme}}{1993}]{vanhamme93}
{van Hamme} W.,  1993, \aj, 106, 2096

\bibitem[\protect\citeauthoryear{{van Winckel}}{{van
  Winckel}}{2003}]{vanwinckel03}
{van Winckel} H.,  2003, \araa, 41, 391

\bibitem[\protect\citeauthoryear{{Van Winckel}, {Waelkens}, {Fernie} \&
  {Waters}}{{Van Winckel} et~al.}{1999}]{vanwinckel99}
{Van Winckel} H.,  {Waelkens} C.,  {Fernie} J.~D.,    {Waters} L.~B.~F.~M.,
  1999, \aap, 343, 202

\bibitem[\protect\citeauthoryear{{Van Winckel}, {Waelkens} \& {Waters}}{{Van
  Winckel} et~al.}{1995}]{vanwinckel95}
{Van Winckel} H.,  {Waelkens} C.,    {Waters} L.~B.~F.~M.,  1995, \aap, 293,
  L25

\bibitem[\protect\citeauthoryear{{Van Winckel}, {Waelkens}, {Waters},
  {Molster}, {Udry} \& {Bakker}}{{Van Winckel} et~al.}{1998}]{vanwinckel98}
{Van Winckel} H.,  {Waelkens} C.,  {Waters} L.~B.~F.~M.,  {Molster} F.~J.,
  {Udry} S.,    {Bakker} E.~J.,  1998, \aap, 336, L17

\bibitem[\protect\citeauthoryear{{Waters}, {Waelkens}, {Mayor} \&
  {Trams}}{{Waters} et~al.}{1993}]{waters93}
{Waters} L.~B.~F.~M.,  {Waelkens} C.,  {Mayor} M.,    {Trams} N.~R.,  1993,
  \aap, 269, 242

\bibitem[\protect\citeauthoryear{{Wilson}}{{Wilson}}{2008}]{wilsonDDE}
{Wilson} R.~E.,  2008, \apj, 672, 575

\bibitem[\protect\citeauthoryear{{Wilson}, {Chochol}, {Kom{\v z}{\'{\i}}k},
  {Van Hamme}, {Pribulla} \& {Volkov}}{{Wilson} et~al.}{2009}]{wilson09}
{Wilson} R.~E.,  {Chochol} D.,  {Kom{\v z}{\'{\i}}k} R.,  {Van Hamme} W.,
  {Pribulla} T.,    {Volkov} I.,  2009, \apj, 702, 403

\bibitem[\protect\citeauthoryear{{Wilson} \& {Devinney}}{{Wilson} \&
  {Devinney}}{1971}]{wd}
{Wilson} R.~E.,  {Devinney} E.~J.,  1971, \apj, 166, 605

\bibitem[\protect\citeauthoryear{{Wood}}{{Wood}}{1990}]{wood90}
{Wood} P.~R.,  1990, in {M.~O.~Mennessier \& A.~Omont} ed., From Miras to
  Planetary Nebulae: Which Path for Stellar Evolution? {Pulsation and evolution
  of Mira variables}.
pp 67--84

\bibitem[\protect\citeauthoryear{{Wood} \& {Arnett}}{{Wood} \&
  {Arnett}}{2011}]{woodarnett}
{Wood} P.~R.,  {Arnett} W.~D.,  2011, in in press ed., Why Galaxies Care About
  AGB Stars II Testing a modified mixing-length theory: comparison to the
  pulsation of agb stars

\bibitem[\protect\citeauthoryear{{Wood} \& {et al. (MACHO
  Collaboration)}}{{Wood} \& {et al. (MACHO Collaboration)}}{1999}]{wood99mn}
{Wood} P.~R.,  {et al. (MACHO Collaboration)} 1999, in {Le Bertre} T.,  {Lebre}
  A.,   {Waelkens} C.,  eds, IAU Symp. 191: Asymptotic Giant Branch Stars
  {MACHO observations of LMC red giants: Mira and semi-regular pulsators, and
  contact and semi-detached binaries}.
Astronomical Society of the Pacific, San Francisco, pp 151--+

\bibitem[\protect\citeauthoryear{{Wood} \& {Nicholls}}{{Wood} \&
  {Nicholls}}{2009}]{dust}
{Wood} P.~R.,  {Nicholls} C.~P.,  2009, \apj, 707, 573

\bibitem[\protect\citeauthoryear{{Wood}, {Olivier} \& {Kawaler}}{{Wood}
  et~al.}{2004}]{sequenceDstars}
{Wood} P.~R.,  {Olivier} E.~A.,    {Kawaler} S.~D.,  2004, \apj, 604, 800

\bibitem[\protect\citeauthoryear{{Wood} \& {Sebo}}{{Wood} \&
  {Sebo}}{1996}]{woodsebo}
{Wood} P.~R.,  {Sebo} K.~M.,  1996, \mnras, 282, 958

\bibitem[\protect\citeauthoryear{{Zahn}}{{Zahn}}{1977}]{zahn}
{Zahn} J.-P.,  1977, \aap, 57, 383

\end{thebibliography}

\label{lastpage}

\end{document}